\documentclass[10pt,a4paper]{article}
\usepackage[utf8]{inputenc}
\usepackage[margin=0.75in]{geometry}
\usepackage[english]{babel}
\usepackage{amsmath}
\usepackage{amsfonts}
\usepackage{amssymb}
\usepackage{graphicx}
\usepackage{braket}
\usepackage{hyperref}
\usepackage{float}
\usepackage[title]{appendix}
\usepackage[backend=bibtex,natbib=true,sorting=none]{biblatex} 

\usepackage[autostyle=true]{csquotes} 
\DeclareMathOperator{\Tr}{Tr}
\DeclareMathOperator{\sech}{sech}
\author{
\normalsize M. Rivera-Tapia$^{1,2^\ast}$, Marcel I. Y\'{a}\~nez-Reyes$^{3,4,\dagger}$, A. Delgado$^{1,2}$, and G. Rubilar$^{2}$
 \\ \vspace{-1mm}
$^1${\small \emph{Instituto Milenio de Investigaci\'on en \'Optica, Universidad de Concepci\'on, Concepci\'on, Chile.}}\\
$^2${\small \emph{Departamento de F\'isica, Facultad Ciencias F\'isicas y Matem\'aticas, Universidad de Concepci\'on,  Concepci\'on, Chile}}\\
$^3${\small \emph{ITFA, University of Amsterdam, Science Park 904, 1018 XE,  Amsterdam,  The Netherlands}}\\
$^4${\small \emph{Nikhef Theory Group, Science Park 105, 1098 XG Amsterdam, The Netherlands}}\\
{\small $^\ast$\texttt{mriverat@udec.cl}, $^{\dagger}$\texttt{marcelyr@nikhef.nl}}
}
\date{}

\title{Outperforming classical estimation of Post-Newtonian parameters of Earth's gravitational field using quantum metrology} 
\begin{document}
\maketitle
\begin{abstract}
The Hong-Ou-Mandel (HOM) effect is analyzed for photons in a modified Mach-Zehnder setup with two particles experiencing different gravitational potentials, which are later recombined using a beam-splitter. It is found that the HOM effect depends directly on the relativistic time dilation between the arms of the setup. This temporal dilation can be used to estimate the $\gamma$ and $\beta$ parameters of the parameterized post-Newtonian formalism. The uncertainty in the parameters $\gamma$ and $\beta$ are  of the order $ 10^{-8}-10^{-12}$, depending on the quantum state employed. 
\end{abstract}

\section{Introduction} \label{SEC1}
General Relativity is a non-linear and metric theory of the gravitational field. The non-linear character of General Relativity has as consequence that very few analytical solutions are known most of which require a high degree of symmetry. In order to compare the predictions of the theory with observational data certain approximations have become invaluable tools. For instance, gravitational waves are often described by solutions to linear approximations of the Einstein equations and do not arise as solutions of the exact equations \cite{Auger-2017-GW-book}. Another very useful approximation of General Relativity is the so called post-Newtonian approximation, a method for solving Einstein's field equations for sources that move slowly compared to the speed of light and that generate weak gravitational fields\cite{Will-2011-effectiveness-of-PN-approximation, will-review-2014-confrontation-GR-and-experiment}. This approximation has been intensively employed as a tool to interpret experimental tests of General Relativity and to test alternative metric theories of gravity\cite{Merkowitz-2010-Tests-of-Gravity-Using-Lunar-Laser-Ranging,Zych-2012-GR-effects-in-quantum-interference-of-photons, Clifford-2014-review-on-experimental-tests-of-GR, Clifton-2008-PPN-for-fR-theories}. Furthermore, this approximation is considered to be sufficiently precise for explaining most solar-system tests that will be carried out in the foreseeable future \cite{will-review-2014-confrontation-GR-and-experiment}. 
The Hong-Ou-Mandel (HOM) effect is a purely quantum effect with no classical counterpart. This effect is observed in an optical arrangement consisting of two photons arriving at a beam-splitter. Classically, the photons should exit the system and be detected at two different ports. Nevertheless, the quantum mechanical description allows for the possibility that the photons emerge at the same output port or at different output ports, depending on whether there are differences in the length of the arms of the array \cite{HOM-original-paper-1987}. In the HOM effect we are interested in the measurement of the coincidence probability, that is, the event in which a single photon is detected at each output port. Several applications of HOM effect have been proposed, such as, for instance, measuring the difference of arrival times between photons, that is, a time sensor \cite{Lyons-2018-attosecond-resolution-HOM-interferometry}, and applications to quantum information tasks, such as implementation of Bell measurements \cite{Pan-2012-Multiphoton-entanglement-and-interferometry} and the generation of high-dimensional entangled states \cite{Zhang-2016-high-dimensional-states-through-quantum-interference, Ndagano-2019-Entanglement-distillation-by-HOM-interference-with-orbital-angular-momentum-states}. Recently, a large improvement on the measurement resolution of arrival times has been reported \cite{Lyons-2018-attosecond-resolution-HOM-interferometry, Chen-2019-HOM-on-a-biphoton-beat-note}.

Modern studies concerning quantum information techniques in relativistic contexts have been performed by Fuentes {\it et al.} \cite{Fuentes-Entanglement-in-non-inertial-frames}, where acceleration effects on quantum entanglement in massless two-mode quantum fields were studied. Refinements using massive particles and particles with spin have been carried out \cite{Montero-2011-Entanglement-in-non-inertial-frames-with-spin}. Furthermore, these predictions have already been experimentally confirmed \cite{Fink-2017-Experimental-test-photonic-entanglement-in-non-inertial-frames}. On the other hand, recent studies focused instead in effects induced by gravity in large-scale optical arrays \cite{Brodutch-2015-PN-effects-in-optical-interferometry,Maldito-Brady-2020-HOM-in-relativistic-setup,Terno-2020-Large-scale-optical-interferometry-in-general-spacetimes}, developing potential applications of quantum entanglement in general relativity such as the estimation of the gravitational acceleration utilizing photons \cite{MALDITOCHEN-2019-ESTIMATION-GRAVITATIONAL-ACCELERATION-QUANTUM-OPTICAL-INTERFEROMETERS}, the usage of optical arrays in the post-Newtonian formalism \cite{Brodutch-2015-PN-effects-in-optical-interferometry,Maldito-Brady-2020-HOM-in-relativistic-setup,Terno-2020-Large-scale-optical-interferometry-in-general-spacetimes}, the implementation of Sagnac arrays to measure a rotation parameter of G\"odel metric \cite{Delgado_2002-Quantum-Gyroscopes-and-godels-universe,Delgado-2006-Sagnac-in-godels-universe}, the usage of optical arrays to generate a path-entangled state induced by gravitational time delay in photons \cite{autocita}, and the proposal of a HOM array to study the effect of the gravitational drag \cite{Maldito-Brady-2020-HOM-in-relativistic-setup}. Furthermore, quantum metrology has been considered as a tool for  the estimation of gravitational parameters. Kohlrus {\it et al.} studied how photons propagating between satellites in a Kerr spacetime can be used to estimate the equatorial velocity and Schwarzschild radius of the gravitational source. In a further study, the same group showed how the rotation in the photon's polarization induced by the Kerr metric can be utilized to estimate values of Earth's radius and the distance between satellites \cite{Fuentes-2019-Quantum-metrology-estimation-of-spacetime-parameters-of-earth-outperforming-classical-precision}. Finally, Kish {\it et al.} studied the Kerr rotation parameter by using a Mach-Zehnder interferometer \cite{Kish-2019-Quantum-metrology-in-kerr}. 

 Motivated by studies of quantum metrology applied in estimation of spacetime parameters and taking into account the need for enhancing parameter estimation with respect to classical estimations in General Relativity, we study the quantum estimation of post-Newtonian parameters using a HOM array. We show how the HOM effect is produced by a path difference between the arms of the interferometer induced by the gravitational field. This effect depends on the gravitational time dilation between the arrival of photons to the detectors, and the wavepacket dispersion. We use this effect to estimate the arrival times in order to calculate the bounds for the precision of the estimation of the post-Newtonian parameters. In order to compare the quantum and classical bounds for the precision we utilize a scheme which only measures arrival times and we compare it with a scheme that relies on the HOM effect. To improve the estimation using the HOM effect, we consider separable and two-mode squeezing vacuum states. Finally, we compare our results with current values and discuss the implications and improvements of applying quantum metrology to this problem for the first time.

This article is organized as follows: in Sec.~\ref{SEC2} we compute the temporal delay in a HOM array. In Sec.~\ref{SEC3} we calculate the HOM effect induced by the gravitational time dilation. In Sec.~\ref{SEC4} we estimate the post-Newtonian parameters using the HOM array. In Sec.~\ref{SEC5} we calculate the uncertainties in the measurement of the parameters. Finally, in Sec.~\ref{SEC6} we summarize our results and conclude.  

\section{Temporal delay in a Hong-Ou-Mandel array} \label{SEC2}
In this section we describe the weak field approximation and the metric utilized to calculate the temporal delays of photons that travel in a Hong-Ou-Mandel array. 

The post-Newtonian (PN) formalism is a framework that allows us to describe the effects of the gravitational field using a perturbative expansion, which in its simpler form, depends on the Newtonian gravitational potential. We shall use the following spacetime metric in isotropic coordinates within the PN formalism, 
      \begin{eqnarray}
      ds^2 = \left( 1 +2 \frac{\phi(x)}{c^2} + 2\beta \frac{\phi^{2}(x)}{c^4} \right)c^2 dt^2 - \left(1-2\gamma \frac{\phi(x)}{c^2}\right) \delta_{ij} dx^{i}dx^{j},
      \end{eqnarray}
where $\gamma$ is a constant that parametrizes  how much spatial curvature is generated by the gravitating mass, and $\beta$ characterizes the non-lineal contribution to the metric field. Our main goal is to use the HOM effect to determine deviations of both parameters from the values within Einstein's theory,  $\gamma = \beta =1$.  

\begin{figure}
    \centering
    
        \includegraphics[width=0.9\textwidth]{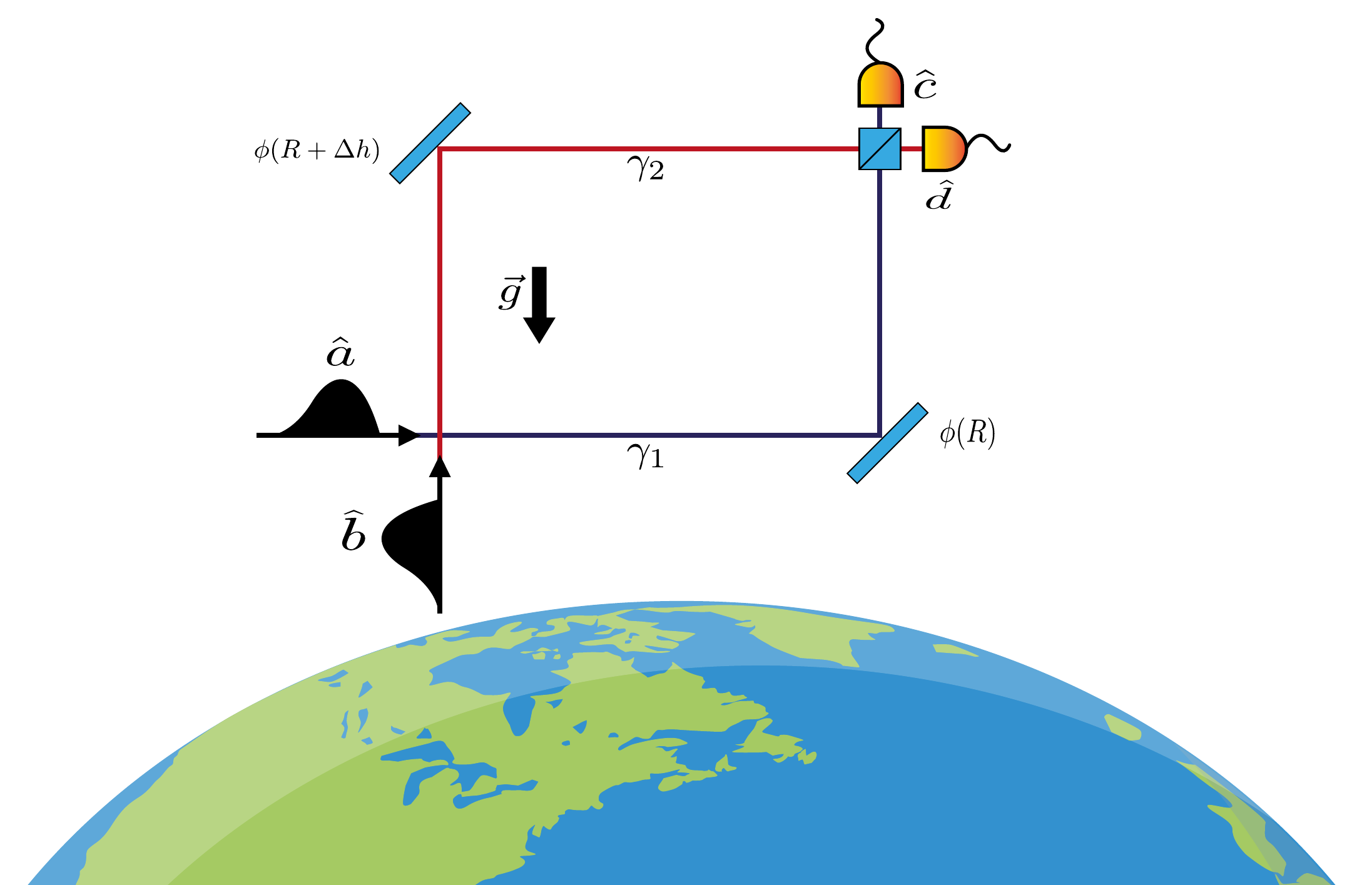} 
          \caption{ HOM interferometer: two photons are injected to the array. Each photon follows  different paths. Each photon experiences a different gravitational potential, for photons moving through horizontal part of path $\gamma_1$ experience a gravitational potential $\phi(R)$, and for photons moving along horizontal part of path $\gamma_2$  experience a potential $\phi(R+\Delta h)$. Finally photons are recombined at the beam-splitter localized at that potential. The temporal delay in the arrival time of photon is measured by a clock localized besides the detectors.}
          \label{HOM}
\end{figure}

 Let us consider a HOM interferometer with light sources located at a height $R$, detectors at $R+\Delta h$ and two paths/arms $(\gamma_2,\gamma_1)$, as  shown in \ref{HOM}. We use the defining property of photons, that is, the fact they move along  light-like curves ($ds^2=0$), which implies that the change in temporal coordinates is given by
       \begin{eqnarray}
       \Delta t &=& \frac{1}{c}\int dx \left( 1 +2\frac{\phi}{c^2} + 2\beta \frac{\phi^2}{c^4} \right)^{-1/2} \left(1-2\gamma \frac{\phi}{c^2} \right)^{1/2} \nonumber \\
       &\approx & \frac{1}{c} \int dx \left( 1 +\frac{\phi}{c^2} + (\frac{3}{2}-\beta) \frac{\phi^2}{c^4} \right) \left(1 -\gamma \frac{\phi}{c^2} - \frac{\gamma^2}{2}\frac{\phi^2}{c^4} \right). \label{DeltaT}
       \end{eqnarray}
    Furthermore, the proper length of an arbitrary interval in this metric is given by
     \begin{eqnarray}
    L &=& \int dx \left( 1 -2 \gamma \frac{\phi}{c^2} \right)^{1/2} \nonumber \\
    &\approx & \int dx \left( 1 -\gamma \frac{\phi}{c^2}  -\frac{\gamma^2}{2}\frac{\phi^2}{c^4} \right).
	\end{eqnarray}       
    Hence, in the HOM interferometer, the proper length of each arm is 
      \begin{eqnarray}
      L_{\gamma_1}& \approx & \left( 1 - \gamma \frac{\phi(R)}{c^2}-\frac{\gamma^2 \phi^2(R)}{2c^4} \right) \Delta x_{\gamma_1}, \label{ProperLgamma1}\\
      L_{\gamma_2}& \approx & \left( 1 - \gamma \frac{\phi(R+\Delta h)}{c^2}-\frac{\gamma^2 \phi^2(R+ \Delta h)}{2c^4} \right) \Delta x_{\gamma_1},
      \label{ProperLgamma2}
      \end{eqnarray}
where $\Delta x_{\gamma_1}$ is the interval of spatial coordinates in the horizontal path.    
       
For each one of the paths $\gamma_1$ and $\gamma_2$ in the interferometer we have that the interval of temporal coordinates (\ref{DeltaT}) in terms of the proper lengths (\ref{ProperLgamma1}) and (\ref{ProperLgamma2}) become
	\begin{eqnarray}
	\Delta t_{\gamma_2} &\approx & ( 1 - \frac{\phi(R+\Delta h)}{c^2} + (\frac{3}{2} -\beta) \frac{\phi^2(R+\Delta h)}{c^4})\frac{L_{\gamma_2}}{c},
	\label{dt1}
    \end{eqnarray}	      
       and
       \begin{eqnarray}
       \Delta t_{\gamma_1} \approx ( 1 -  \frac{\phi(R)}{c^2} + (\frac{3}{2} -\beta) \frac{\phi^2(R)}{c^4})\frac{L_{\gamma_2}}{c}.
	\label{dt2}
       \end{eqnarray}
  
In order to observe interference, we demand the condition $\Delta x_{\gamma_1} = \Delta x_{\gamma_2}$, that is, an array placed on a equipotencial surface is balanced and there is no difference of optical length between the arms of the array. With this constraint, the proper length $L_{\gamma_1}$, to second second order in $\phi/c^2$, can be written in terms of $L_{\gamma_2}$ as
  \begin{eqnarray}
  \frac{L_{\gamma_1}}{L_{\gamma_2}} 
  & \approx & 1+ \gamma \frac{g\Delta h}{c^2} + 2\gamma^2 \frac{g\Delta h \phi(R)}{c^4}.
\end{eqnarray} 
Then, the proper length of $\gamma_1$ becomes
\begin{eqnarray}
L_{\gamma_1} \approx \left( 1+ \gamma \frac{g\Delta h}{c^2} + 2\gamma^2 \frac{g\Delta h \phi(R)}{c^4} \right)L_{\gamma_2}. 
\end{eqnarray}       
Moreover, proper time measured by a clock located at the detectors, which are  at the gravitational potential $\phi(R+\Delta h)$, is given by
\begin{eqnarray}
\Delta \tau & = & \left( 1 +2 \frac{\phi}{c^2} + 2\beta \frac{\phi^2}{c^4} \right)^{1/2} \Delta t,
\end{eqnarray} 
where $\Delta t = \Delta t^{\rm H}_{\gamma_1} - \Delta t^{\rm H}_{\gamma_2}$ is the difference of temporal coordinates between paths $\gamma_1$ and $\gamma_2$. Thus, using Eqs.~(\ref{dt1}) and (\ref{dt2}) we have
\begin{eqnarray}
\Delta \tau &=& \left( (\gamma +1 )\frac{g\Delta h}{c^2} + 2(\gamma^2 -1 + \beta) \frac{\phi(R)g\Delta h}{c^4}\right) \frac{L_{\gamma_2}}{c}. \label{TimeDelay1}
\end{eqnarray}
 This expression can be recast considering the effective area $A=L_{\gamma_2}\times \Delta h$ of the interferometer. We obtain the expression
 \begin{eqnarray}
\Delta \tau &=& \left( (\gamma +1 ) + 2(\gamma^2 -1 + \beta) \frac{\phi(R)}{c}\right) \frac{A g}{c^3}. \label{TimeDelay2} 
\end{eqnarray}
As we can see from Eq.\thinspace(\ref{TimeDelay2}), this depends on the PN parameters $\gamma$ and $\beta$, where the parameter $\gamma$ is only present to first order in the the gravitational potential and we have contributions of $\gamma$ and $\beta$ to second order in the gravitational potential. According to Eq.~(\ref{TimeDelay2}), given a measurement of the temporal delay $\Delta \tau$ the considered PN parameters are not independent.    

\section{HOM effect induced by gravitational time dilation}\label{SEC3}
Gravitational fields shift the phase of quantum states \cite{Brodutch-2015-PN-effects-in-optical-interferometry}. This shift is accumulated through different paths of the interferometer and is the source of interference in our system. To calculate the probability of detection in the interferometer we incorporate the effect of gravity by using a unitary operator $U(\varphi)$, where $\varphi$ is the phase. As we know, if we are working in regime where quantum gravity effects are not relevant \cite{shore-2003-quantum-theory-of-light-propagation-in-GR}, the propagation of an electromagnetic field is described by the classical wave equations on a curved background. The usual approach consists in describing the propagation  of an electromagnetic wave by means of geometric optics \cite{Will-2018-Book-Theory-and-experiment-in-gravitational-physics}. In this approximation, the phase of the vector potential satisfies the eikonal equation, which corresponds to the Hamilton-Jacobi equation for massless particles. In the case of an interferometric experiment, and in a static spacetime, the gravitationally induced phase shift is given by $\Delta \Psi = \bar{\omega} \Delta t$, where $\bar{\omega}$ is the frequency measured by an observer at the infinity and $\Delta t$ is the interval of difference of temporal coordinate between paths of the array. The phase shift can be recast, considering the frequency $\omega$ and the proper time $\Delta \tau$ measured by an observer experiencing a specific gravitation potential, as $\Delta \Psi = \omega \Delta \tau$, with $\omega = \bar{\omega}/\sqrt{g_{00}}$ and $\Delta\tau = \sqrt{g_{00}}$. 
\subsection{Two-photon separable state}
Let us consider an initial two-photon wave packet of the form
 \begin{equation}
\ket{\Psi}_{\rm in} = \int d\omega_1 \phi(\omega_1) \hat{v}^{\dagger}(\omega_1)\int d\omega_2 \phi(\omega_2) \hat{u}^{\dagger}(\omega_2)\ket{0}_{12},
\label{Qstate1}
\end{equation}
where $\hat{v}^{\dagger}$  creates a photon with frequency $\omega_1$, and  $ \hat{u}^{\dagger}$ creates a photon with frequency $\omega_2$. The state that describes the photons after the phase shift produced by the gravitational field and the recombination produced by the beam-splitter is
\begin{eqnarray}
\ket{\Psi}_{\rm out} &=& \frac{1}{2}\iint f(\omega_1,\omega_2) e^{-i(\omega_1 \Delta \tau_{\gamma_1} + \omega_2 \Delta \tau_{\gamma_2} + \phi)} \left[i a^{\dagger}(\omega_1) + b^{\dagger}(\omega_1) \right]\left[ a^{\dagger}(\omega_1) + i b^{\dagger}(\omega_1) \right] \vert 0 \rangle_{12}, \nonumber \\
\end{eqnarray}
where $a, a^\dagger, b$ and $b^\dagger$ are creation and annihilation operators of modes at the output ports of the beam-splitter.

Hence, using Born's rule, the probability of simultaneous detection is given by
\begin{eqnarray}
 p_{cd} &=& \frac{1}{4}\iint d\omega_1 d\omega_2 \vert \phi(\omega_1,\omega_2)  \vert^2 \left[ 2  - e^{i \left(\omega_1 - \omega_2 \right) \tau_{\gamma_2}} e^{-i \left(\omega_1 - \omega_2 \right) \tau_{\gamma_1}} -   e^{-i \left(\omega_1 - \omega_2 \right) \tau_{\gamma_2}} e^{i \left(\omega_1 - \omega_2 \right) \tau_{\gamma_1}} \right]. \nonumber \\
\end{eqnarray}

If the photons have Gaussian frequency distributions, with possibly different spectral distributions $\sigma_1$ and $\sigma_2$ and mean frequencies $\omega_1$ and $\omega_2$, the detection probability becomes 
\begin{eqnarray}
p_{cd} &=& \frac{1}{2}\left( 1 - \cos((\omega_1 - \omega_2)\Delta \tau)e^{-\frac{\Delta \tau (\sigma^{2}_1 + \sigma^{2}_2)}{4}} \right),
\label{ProbCoincGeneral}
\end{eqnarray}
where $\Delta \tau$ is the difference of arrival time measured by a clock placed at the detectors. In the case of photons with the same spectral distributions $\sigma_1 = \sigma_2 = \sigma$ and equal mean frequencies $\omega_1 = \omega_2$, the probability of coincidence detection Eq.~\thinspace(\ref{ProbCoincGeneral}) becomes
  \begin{eqnarray}
    p_{cd} &=& \left(1-e^{-\sigma^2 \Delta \tau^2/2} \right)/2. \label{ProbCoinc}
  \end{eqnarray} 
According to Eq.\thinspace(\ref{ProbCoincGeneral}) the coincidence detection probability decreases exponentially as the temporal delay between photons increases. Moreover, the coincidence detection probability also exhibits an harmonic oscillation given by a cosine function, which depends on the temporal delay times the difference between the mean frequency of the photons. This last feature vanishes if we have twin photons with the same frequency, and only the exponential decrease of the probability is present, according to Eq.~\thinspace(\ref{ProbCoinc}). The HOM effect is produced here by the gravitational time dilation, even if both arms have the same proper lengths. If we consider the array placed on a gravitational equipotential surface, there is no difference of proper time, both arms have the same proper length and consequently the coincidence detection probability given by Eq.~\thinspace(\ref{ProbCoinc}) vanishes.	 

\begin{figure}
\centering
        \includegraphics[width=0.8\textwidth]{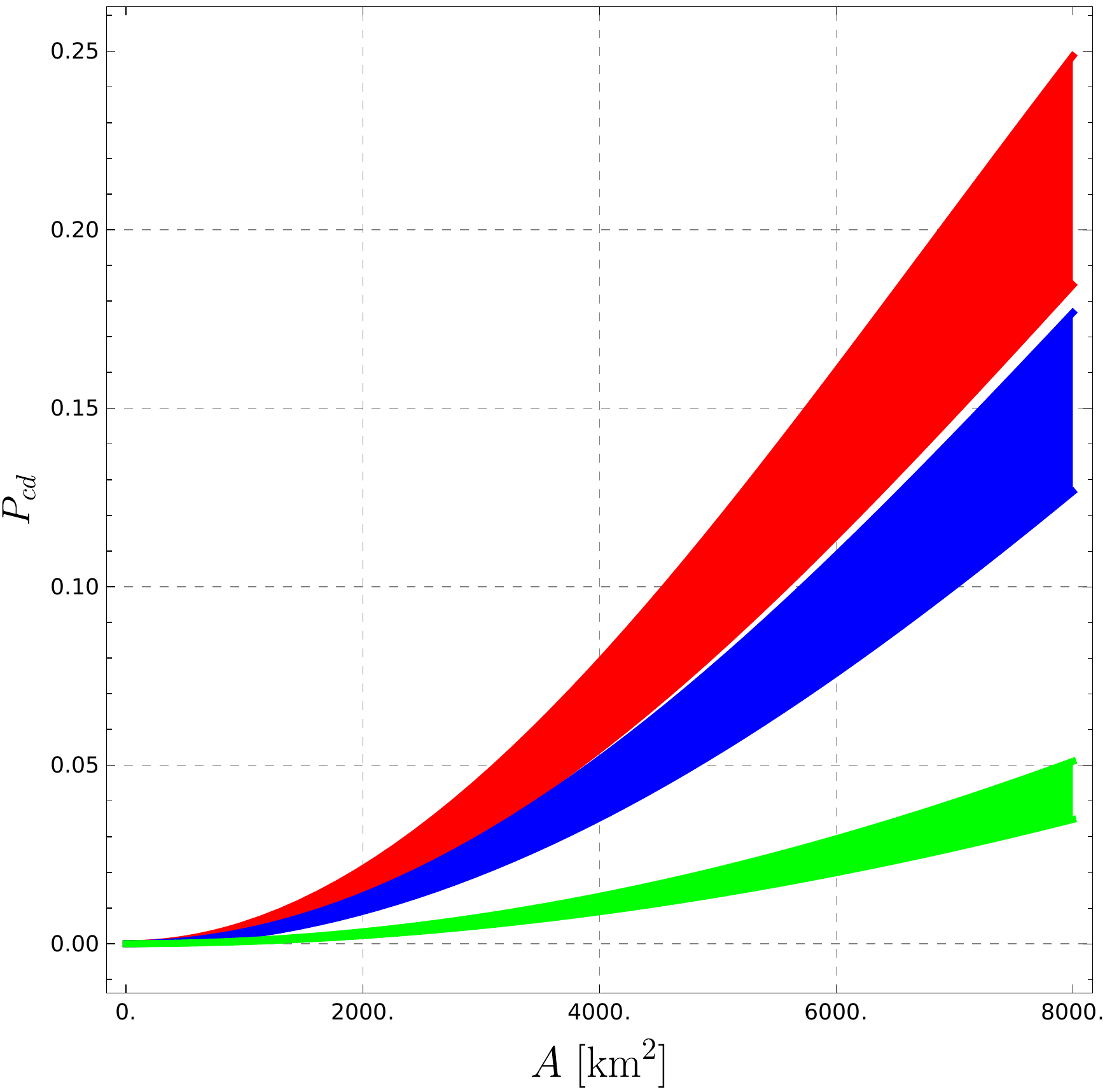} 
        \caption{ Probability of detection in coincidence for HOM interferometer according to Eq.\thinspace(\ref{ProbCoinc}) for different values of $\gamma$ and $\beta$ in the interval $[0.8,1.2]$ and different wavelength width. For each color, upper  line represents $\beta = \gamma =1.2$ and mean wavelength $\lambda = 3.3\;[\rm \mu m]$, and lower line $\beta=\gamma= 0.8$. Red lines $\delta \lambda = 0.370\;[\mu m]$, blue lines $\delta \lambda =0.296 \;[\mu m]$ and green lines $\delta \lambda =0.148 \;[\mu m]$ .}
        \label{PHOM}
\end{figure}

\begin{figure}[H]
\centering
        \includegraphics[width=0.8\textwidth]{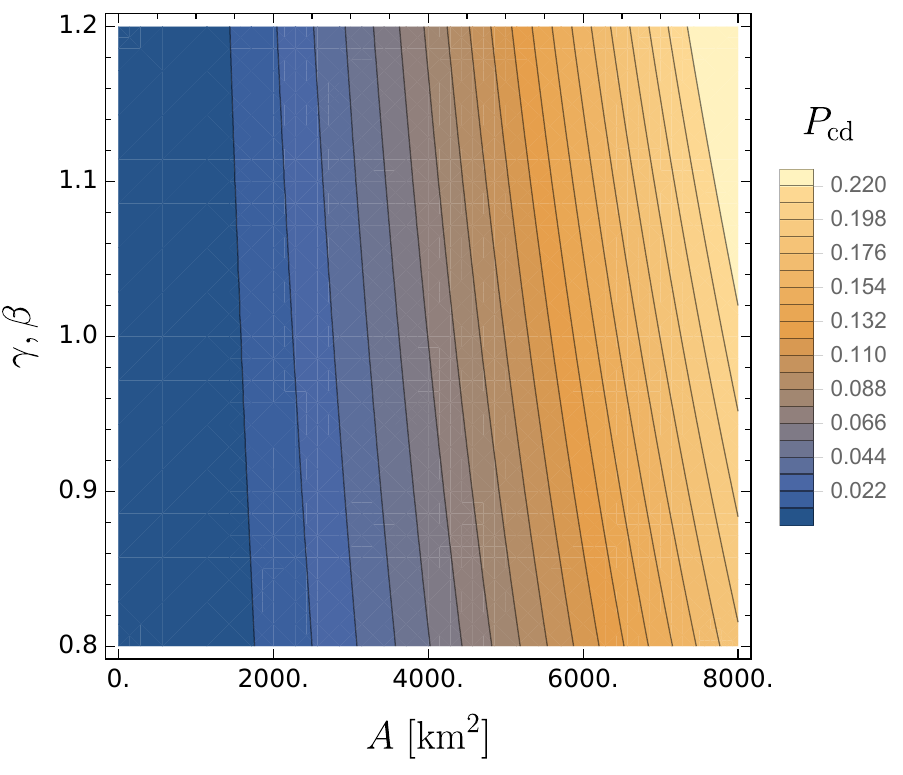} 
        \caption{ Contour plot for the probability of detection in coincidence for HOM interferometer according to Eq.\thinspace(\ref{ProbCoinc}) in terms of different values of $\gamma$ and $\beta$ in the interval $[0.8,1.2]$ and the area $A=L_{\gamma_2}\times \Delta h$. The mean wavelength is $\lambda = 3.3\;[\rm \mu m]$, and $\delta \lambda = 0.370\;[\mu m]$.}
        \label{CP-PHOM}
\end{figure}

 Fig.~\ref{PHOM} shows behavior of the coincidence detection probability according to Eq.~(\ref{ProbCoinc}) for different values of the parameters $\beta$ and $\gamma$. The range of values for $\gamma$ and $\beta$ is selected between 0.8 and 1.2 in order to visualize the sensitivity of the HOM array to small deviations of these parameters with respect to prediction of General Relativity. In Fig.~\ref{CP-PHOM} we plot different values of the parameters $\gamma$ and $\beta$ in terms of the effective area of the HOM interferometer, considering small deviations of actual values of the PN parameters \cite{will-review-2014-confrontation-GR-and-experiment}. As  is apparent from this figure, the coincidence detection probability of Eq.~(\ref{ProbCoinc}) has nearly the same variation for each value of $\gamma$ and $\beta$ in the case of both parameter with the same values. 

To take advantage of the HOM effect, we measure the coincidence detection probability given by the operator
\begin{eqnarray}
\hat{S} &=& \int d\omega a^{\dagger}(\omega) \vert 0 \rangle \langle 0 \vert a(\omega) \otimes \int d\omega b^{\dagger}(\omega) \vert 0 \rangle \langle 0 \vert b(\omega). \label{CoincOperator}
\end{eqnarray}
This is such that, 
\begin{eqnarray}
\langle \hat{S}^2 \rangle = \langle \hat{S} \rangle.
\end{eqnarray}
 Therefore, 
\begin{eqnarray}
\langle \hat{S} \rangle &=& \frac{1}{2}\left( 1 - e^{-\sigma^2 \Delta \tau^2/2} \right),\label{ProbObs}
\end{eqnarray}
and
\begin{eqnarray}
\langle (\Delta \hat{S})^2 \rangle &=& \frac{1}{4}\left( 1 - e^{-\sigma^2 \Delta \tau^2} \right).
\label{meansquareHOM}
\end{eqnarray}
Considering Eqs.~(\ref{ProbObs}) and (\ref{meansquareHOM}) we can calculate the root mean square of the coincidence operator in Eq.~(\ref{CoincOperator}), where as the elapsed proper time increases, the dispersion present in the observable increases. It is interesting to note that considering Eq.~(\ref{ProbObs}) we can obtain an estimation of the temporal delay $\Delta \tau$ through the mean value of the coincidence operator $\hat S$.

\subsection{Two-mode squeezed-vacuum state}
Let us assume that the HOM array is fed a two-mode squeezed-vacuum state, which in the Fock basis is given by 
\begin{eqnarray}
\vert \zeta \rangle &=& \frac{1}{\cosh(r)}\sum_{n=0}^{\infty}(-1)^n e^{in\theta}\left( \tanh(r)\right)^n \vert n, n \rangle.
\label{SingleVacuumState}
\end{eqnarray}
Here, $\theta$ is a controllable phase for the squeezed vacuum, $r$ is the squeezing parameter of the two-mode squeezer, and $n$ labels Fock states. The probability of detection (for further details see appendix \ref{ApSingleVacuumState}) of the two photons in coincidence is given by $p_{\rm cd} = \Tr\left( \rho_{\rm squeezing} \hat{S} \right)$, where the operator $\hat{S}$ is given by Eq.\thinspace(\ref{CoincOperator}) and $\rho_{\rm squeezing}=|\zeta\rangle\langle\zeta|$. The probability of coincidence detection in the case of different frequencies, that is, $\omega_1 \neq \omega_2$, becomes 
\begin{eqnarray}
\langle S \rangle &=& \frac{1}{2}\left(\frac{\tanh(r)}{\cosh(r)} \right)^2  \left( 1 - \cos((\omega_1 - \omega_2)\Delta \tau)e^{-\frac{\Delta \tau (\sigma^{2}_1 + \sigma^{2}_2)}{4}} \right).
\label{SingleVacuumStateProbabilityGeneral}
\end{eqnarray}
If the mean frequencies of the Gaussian wave packets are equal, then we obtain
\begin{eqnarray}
\langle S \rangle &=& \frac{1}{2} \left(\frac{\tanh(r)}{\cosh(r)} \right)^2  \left( 1 - \exp\left( -\sigma^2 \Delta \tau^2/2\right) \right).
\label{SingleVacuumStateProbability}
\end{eqnarray}
We can recast Eqs.~(\ref{SingleVacuumStateProbabilityGeneral}) and (\ref{SingleVacuumStateProbability}) considering Eqs.~(\ref{ProbCoincGeneral}) and (\ref{ProbCoinc}), respectively. In this case, we finally have that the expectation value is given by
\begin{eqnarray}
\langle S \rangle &=& \left(\frac{\tanh(r)}{\cosh(r)} \right)^2 P_{cd}.
\end{eqnarray}
Thus, the coincidence detection probability of two-mode squeezed-vacuum state is equal to the coincidence detection probability for a two-photo state times a function that depends on the squeezing parameter $r$. In the limit where $r \rightarrow 0$, that is, the squeezing parameter vanishes, the coincidence detection probabilities Eqs.~(\ref{SingleVacuumStateProbability}) and (\ref{SingleVacuumStateProbabilityGeneral}) become Eqs.~(\ref{ProbCoinc}) and (\ref{ProbCoincGeneral}), respectively.

\section{Estimation of $\gamma$ and $\beta$ parameters} \label{SEC4}
\subsection{Parameter estimation by an external clock}
\begin{figure}
    \centering
        \includegraphics[width=1.0\textwidth]{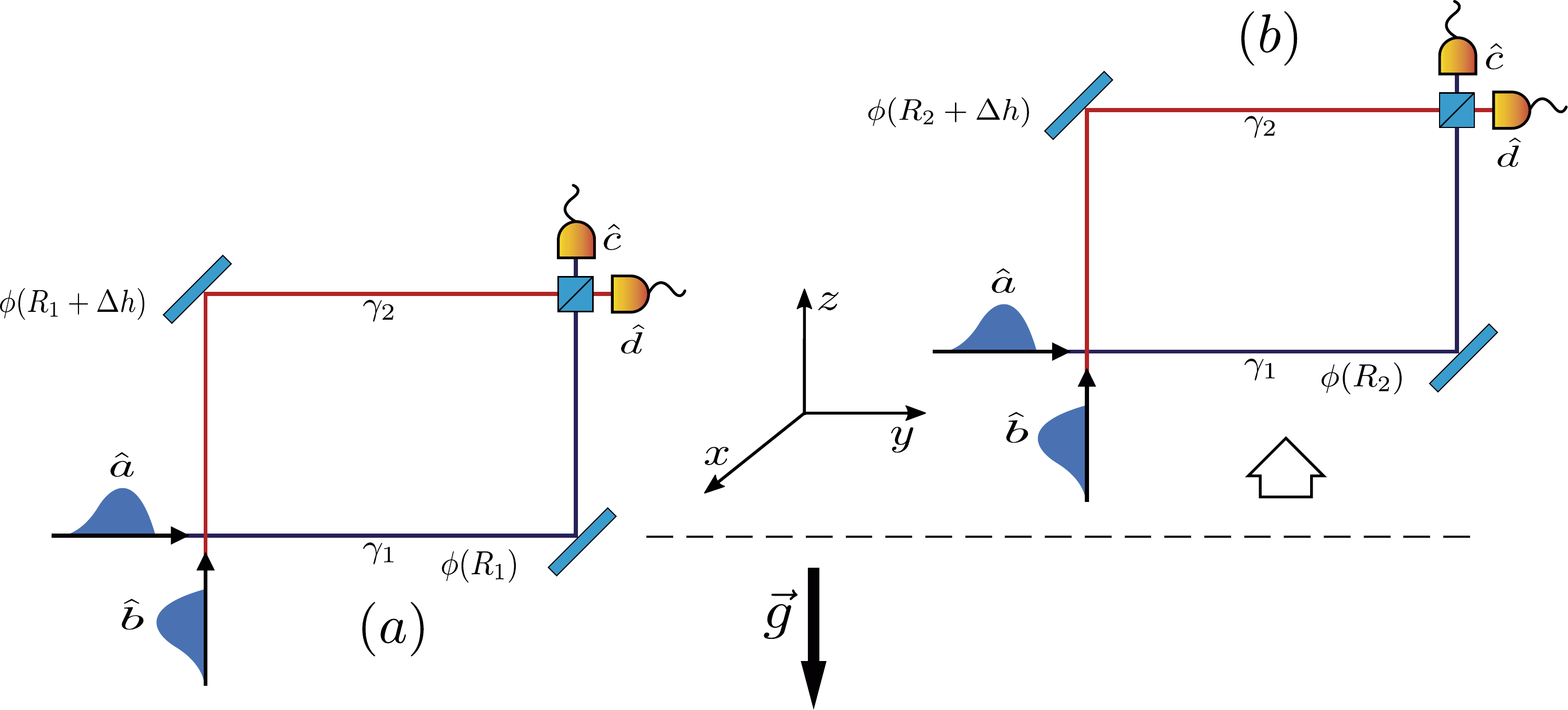} 
          \caption{ HOM interferometer configuration: in order to obtain the values of the PN parameters $\gamma$ and $\beta$ we measure two different times $\Delta \tau_1$ (\ref{DeltaTau1}) and $\Delta \tau_2$ (\ref{DeltaTau2}) (or the probability of detection in coincidence  (\ref{ProbCoinc}) for each arrival time) for each configuration (a) and (b), respectively. In (a) two photons enter to HOM interferometer, bottom path is in a potential $\phi(R_1)$ and the upper path in $\phi(R_1 + \Delta h)$. Both photons are recombined at this potential. For (b) all array is moved a distance such that both photons enter to the array at $\phi(R_2)$, and finally they are recombined at $\phi(R_2 + \Delta h)$.}
          \label{HOM2}
\end{figure}

Considering the temporal delay given by Eq.~(\ref{TimeDelay1}) we use a quantum protocol to obtain an estimate of the parameters $\gamma$ and $\beta$. We consider two different arrival times, for each one we set the HOM interferometer with the lower arm experiencing a gravitational potential $\phi(R_i)$, with $i=1,2$, and the upper arm experiencing the potential $\phi(R_i + \Delta h)$ such that the difference of gravitational potential between each arm (for both interferometers) is (at first order) $g \Delta h$ and each upper arm of the different arrays has a proper length $L_{\gamma_2}$ and $L_{\gamma'_2}$, respectively.  A classical approach consists in measuring the time delay of a light wave-packet in the HOM interferometer, but a quantum approach would be based on measuring the difference of proper time through the probability of detection in the HOM array. Having an estimate of the temporal delays in each array, we solve the following system of equations
\begin{eqnarray}
\Delta \tau_1 &=& \left( (\gamma +1 )\frac{g\Delta h}{c^2} + 2(\gamma^2 -1 + \beta) \frac{\phi(R_1)g\Delta h}{c^4}\right) \frac{L_{\gamma_2}}{c} , \label{DeltaTau1} \\
\Delta \tau_2 &=& \left( (\gamma +1 )\frac{g\Delta h}{c^2} + 2(\gamma^2 -1 + \beta) \frac{\phi(R_2)g\Delta h}{c^4}\right) \frac{L_{\gamma'_2}}{c}. \label{DeltaTau2}
\end{eqnarray}
Where $\Delta \tau_1$ and $\Delta \tau_2$ are the differences of arrival time between each configuration of the array, respectively. $\phi(R_1)$ and $\phi(R_2)$ are the gravitational potential that is experimented by the lower path of each array, and $\Delta h$ is the difference of spatial coordinate in the z direction (see Fig.\thinspace(\ref{HOM2})). Solving for $\gamma$ and $\beta$, we obtain the following solutions
\begin{eqnarray}
\gamma &=& \frac{c^3 \left( L_{\gamma_2}\Delta \tau_2 \phi(R_1) - L_{\gamma'_2} \Delta \tau_1  \phi(R_2) \right)}{g L_{\gamma_2}L_{\gamma'_2}\Delta h \left( \phi(R_1) - \phi(R_2) \right)} -1, \label{gammaDeltaTau}
\end{eqnarray}
and 
\begin{eqnarray}
\beta &=& \frac{2c^3 \left( L_{\gamma_2}\Delta \tau_2 \phi(R_1)-  L_{\gamma'_2} \Delta \tau_1 \phi(R_2)\right)}{L_{\gamma_2} L_{\gamma'_2} g \Delta h \left( \phi(R_1) - \phi(R_2) \right)} + \frac{c^5 \left( L_{\gamma'_2} \Delta \tau_1 - L_{\gamma_2}\Delta \tau_2 \right)}{2 L_{\gamma_2} L_{\gamma'_2} g \Delta h \left(\phi(R_1) - \phi(R_2) \right)} \nonumber \\
&& - \frac{c^6 \left( L_{\gamma_2} \Delta \tau_2 \phi(R_1)- L_{\gamma'_2} \Delta \tau_1 \phi(R_2)\right)^2}{L^{2}_{\gamma_2} L^{2}_{\gamma'_2} g^2 \Delta h^2_1 \left( \phi(R_1) - \phi(R_2) \right)^2}. \label{betaDeltaTau}
\end{eqnarray}
In terms of the areas $A_1$ and $A_2$ of each array the parameters $\gamma$ and $\beta$ become
\begin{eqnarray}
\gamma &=& \frac{c^3 \left(A_1 \Delta \tau_2 \phi(R_1)- A_2 \Delta \tau_1 \phi(R_2) \right)}{g A_1 A_2 (\phi(R_1) - \phi(R_2))}-1,
\label{gammaTauArea}
\end{eqnarray}
\begin{eqnarray}
\beta &=& \frac{2c^3 \left( A_1 \Delta \tau_2 \phi(R_1) - A_2 \Delta \tau_1 \phi(R_2) \right)}{A_1 A_2 g (\phi(R_1) - \phi(R_2))} + \frac{c^5 \left( A_2 \Delta \tau_1 - A_1 \Delta \tau_2\right)}{2 A_1 A_2 (\phi(R_1) - \phi(R_2))} \nonumber \\
&& - \frac{c^6 \left( A_1 \Delta \tau_2 \phi(R_1) - A_2 \Delta \tau_1 \phi(R_2)\right)^2}{A^{2}_1 A^{2}_{2} (\phi(R_1) - \phi(R_2))^2}.
\label{betaTauArea}
\end{eqnarray}
As we can see in the expressions for $\gamma$ and $\beta$, the $\beta$ parameter can be recast in terms of $\gamma$. Thereby, $\beta$ as given by Eq.\thinspace(\ref{betaTauArea}) becomes
\begin{eqnarray}
\beta &=& \frac{c^5 \left( A_2 \Delta \tau_1 - A_1 \Delta \tau_2\right)}{2 A_1 A_2 (\phi(R_1) - \phi(R_2))} + 2 \left( \gamma +1 \right) -\left( \gamma + 1 \right)^2.
\end{eqnarray}
Consequently, our estimates of $\gamma$ and $\beta$ are dependent, because they arise from an estimate of the same observable, which in our case is the elapsed proper time $\Delta \tau$ of Eq.\thinspace(\ref{TimeDelay2}). 

\subsection{Parameter estimation employing a two-photon separable state}

A second way to obtain $\gamma$ and $\beta$ is a slight deviation from the previous one. Instead of considering the time delays of each interferometer (measured with an external clock), we obtain the temporal delay from the probability of detection in the HOM array. Consider the probability of detection in coincidence of Eq.~(\ref{ProbCoinc}), from this probability the time delay (in case of a Gaussian wave-packet) is
\begin{eqnarray}
\Delta \tau_{\rm sep} = \pm \frac{\sqrt{2}}{\sigma}\sqrt{\ln\left( \frac{1}{1-2 \langle S \rangle} \right)},\label{TempDelayProbHOM}
\end{eqnarray}
where $\langle S \rangle$ is the expectation value of the coincidence observable (see Eq.~\ref{ProbObs}), and $\sigma$ is the dispersion of the photons. Replacing Eq.~(\ref{TempDelayProbHOM}) for each HOM interferometer in (\ref{gammaDeltaTau}) and (\ref{betaDeltaTau}) we obtain 
\begin{eqnarray}
\gamma &=& -1 + \frac{c^3 \Delta h}{g L_{\gamma_2}L_{\gamma'_2} \Delta h^2 \left( \phi(R_2)- \phi(R_1) \right)} \nonumber \\
&& \times  \left( \frac{L_{\gamma'_2} \phi(R_1)}{\sigma_1}\sqrt{\ln\left( \frac{1}{1-2 \langle S_1 \rangle} \right)} - \frac{L_{\gamma_2} \phi(R_2)}{\sigma_2}\sqrt{\ln\left(  \frac{1}{1-2 \langle S_2 \rangle} \right)}   \right), \nonumber \\
\label{ParamterGamma}
\end{eqnarray}
where $\sigma_1$, and $\sigma_2$ are the spectral width of the photons employed in the first and second configuration, respectively. $\langle S_1 \rangle$ and $\langle S_2 \rangle$ are the expectation values of the coincidence observable for the first and second configuration, respectively. This expression can be recast as  
\begin{eqnarray}
\gamma &=& -1 + \frac{c^3 \Delta h}{g L_{\gamma_2}L_{\gamma'_2} \Delta h^2 \left( \phi(R_2)- \phi(R_1) \right)} \nonumber \\
&& \times  \left( \frac{L_{\gamma'_2} \phi(R_1)}{\sigma_1}\ln\left( \frac{1}{1-2 \langle S_1 \rangle} \right) - \frac{L_{\gamma_2} \phi(R_2)}{\sigma_2}\ln\left(  \frac{1}{1-2 \langle S_2 \rangle} \right)   \right). \nonumber \\
\label{ParamterGamma2}
\end{eqnarray}
Analogously for $\beta$, we obtain
\begin{eqnarray}
\beta &=& \frac{c^3}{2 g^2 L_{\gamma_2}L_{\gamma'_2} \Delta h^2 \sigma_1 \sigma_2 \left( \phi(R_1) - \phi(R_2)\right)^2} \left[ c^2 g  L_{\gamma_2}L_{\gamma'_2} \Delta h \sigma_1 \sigma_2 \left( L_{\gamma'_2} \sigma_2 \sqrt{\ln\left( \frac{1}{1-2 \langle S_1 \rangle} \right)} \right. \right. \nonumber \\
&& \left. \left. - L_{\gamma_2} \sigma_1 \sqrt{\ln\left( \frac{1}{1-2 \langle S_2 \rangle} \right)} \right) \left( \phi(R_1) - \phi(R_2) \right)  + 4 g L_{\gamma_2} L_{\gamma'_2} \Delta h \sigma_1 \sigma_2 [\phi(R_1) - \phi(R_2)]\right. \nonumber \\
&& \left. \times  \left( L_{\gamma_2} \sigma_1 \sqrt{\ln\left( \frac{1}{1-2 \langle S_1 \rangle} \right)} \phi(R_1)- L_{\gamma'_2} \sigma_2 \sqrt{\ln\left( \frac{1}{1-2 \langle S_2 \rangle} \right)} \phi(R_2)\right) \right. \nonumber \\
&& \left. - 2c^3 \left( L_{\gamma_2} \sigma_1 \sqrt{\ln\left( \frac{1}{1-2 \langle S_1 \rangle} \right)} \phi(R_1)- L_{\gamma'_2} \sigma_2 \sqrt{\ln\left( \frac{1}{1-2 \langle S_2 \rangle} \right)} \phi(R_2)\right) \right].
\label{ParamterBeta}  
\end{eqnarray}

%
%

\subsection{Parameter estimation employing a two-mode squeezed-vacuum state}

In this subsection we employ the probability of detection in coincidence when the HOM array  is injected two single-mode vacuum states. Considering the probability of detection given by the Eq.\thinspace(\ref{SingleVacuumStateProbability}), the temporal delay becomes
\begin{eqnarray}
\Delta \tau_{\rm squeezing} &=&  \left(\frac{\sqrt{2}}{\sigma} \right)\sqrt{\ln\left(\frac{\tanh^2(r)}{2 \langle S \rangle\cosh^2(r) - \tanh^2(r) }\right)},
\label{DeltaTauSqueezing}
\end{eqnarray}
where $\langle S \rangle$ is the probability of detection in coincidence. Replacing the temporal delay Eq.\thinspace(\ref{DeltaTauSqueezing}) the parameter $\gamma$ becomes 
\begin{eqnarray}
\gamma &=& \frac{A_1 c^3 \phi (R_1) \ln \left(\frac{1}{1-2 \langle S_2 \rangle \sech^4(r_2)}\right)-A_2 c^3 \phi (R_2) \ln \left(\frac{1}{1-2 \langle S_1 \rangle \sech^4(r_1)}\right)}{A_1 A_2 g \sigma  \phi (R_1)-A_1 A_2 g \sigma  \phi (R_2)}-1, \nonumber \\
\label{gammaSqueezing}
\end{eqnarray}
where $\langle S_1 \rangle$ and $\langle S_2 \rangle$ are the probabilities of detection in coincidence for the first and second configuration of the HOM array, respectively. $r_1$ and $r_2$ are the expectation values of the squeezing parameter in the first and second configuration, respectively.  For the parameter $\beta$ we obtain
\begin{eqnarray}
\beta &=& -\frac{c^6 \left(A_1 \phi (R_1) \ln \left(\frac{1}{1-2\langle S_2 \rangle \sech^4(R_2)}\right)-A_2 \phi (R_2) \ln \left(\frac{1}{1-2\langle S_1 \rangle \sech^4(R_1)}\right)\right)^2}{A_1^2 A_2^2 g^2 \sigma ^2 \Delta \phi_{12}^2} \nonumber \\
&& +\frac{A_2 c^5 \ln \left(\frac{1}{1-2\langle S_1 \rangle \sech^4(R_1)}\right)-A_1 c^5 \ln \left(\frac{1}{1-2\langle S_2 \rangle \sech^4(R_2)}\right)}{2 A_1 A_2 g \sigma  \phi (R_1)-2 A_1 A_2 g \sigma  \phi (R_2)} \nonumber \\
&& +\frac{2 A_1 c^3 \phi (R_1) \ln \left(\frac{1}{1-2\langle S_2 \rangle \sech^4(R_2)}\right)-2 A_2 c^3 \phi (R_2) \ln \left(\frac{1}{1-2\langle S_1 \rangle \sech^4(R_1)}\right)}{A_1 A_2 g \sigma  \phi (R_1)-A_1 A_2 g \sigma  \phi (R_2)}.
\nonumber \\
\label{betaSqueezing}
\end{eqnarray}

\section{Uncertainties and relative errors of $\gamma$ and $\beta$} \label{SEC5}
In this section, we calculate the uncertainties of the PN parameters $\gamma$ and $\beta$. In the first part we consider errors in the measurement of the arrival time of photons of a classical wave. In the second part we focus on the arrival time obtained from the probability of detection in coincidence, but employing several schemes.

\subsection{Errors in the arrival times}
\begin{figure*}
\centering
\begin{tabular}{c}
\includegraphics[width=90mm]{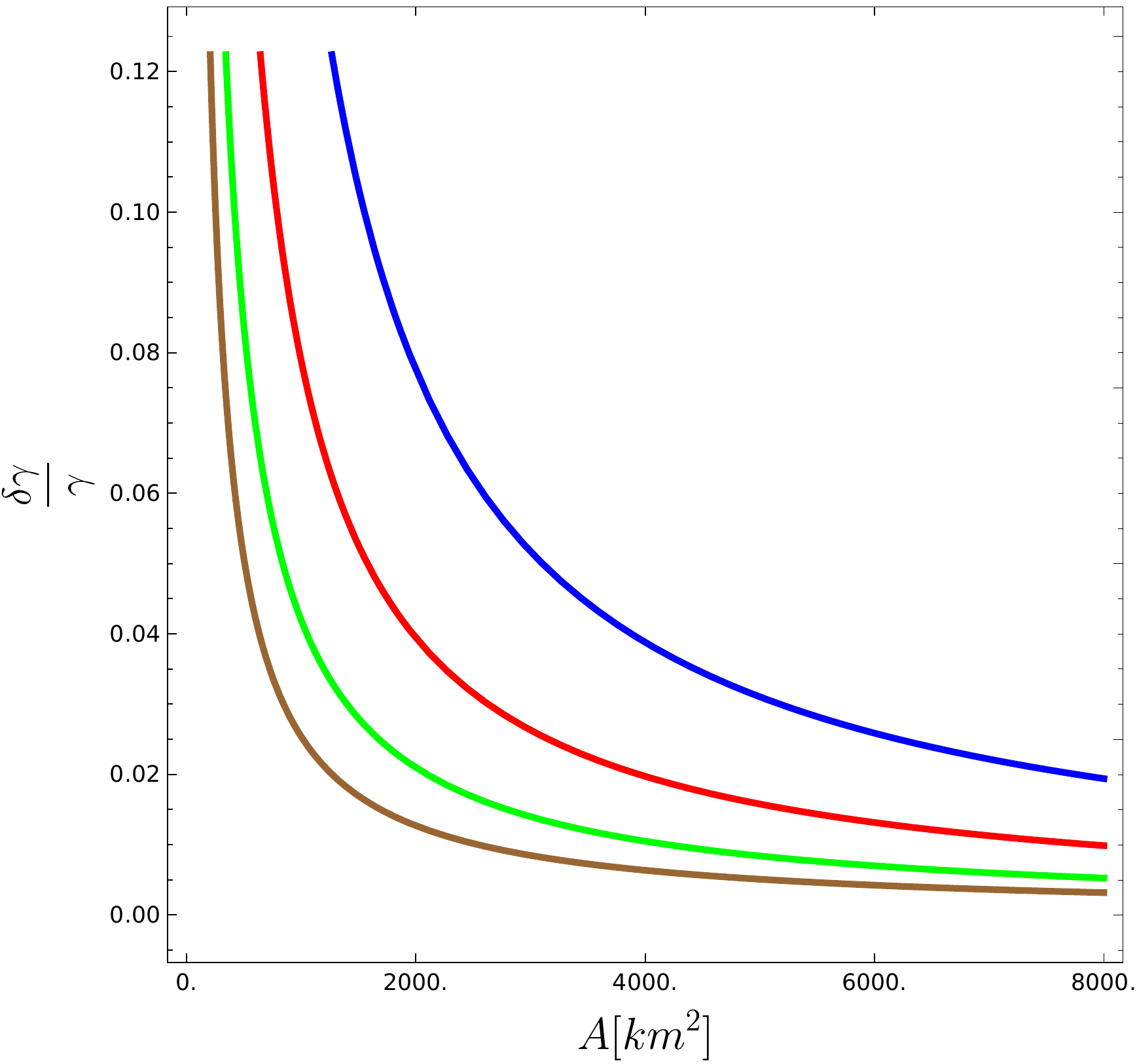} \\
(a) \\
 \includegraphics[width=90mm]{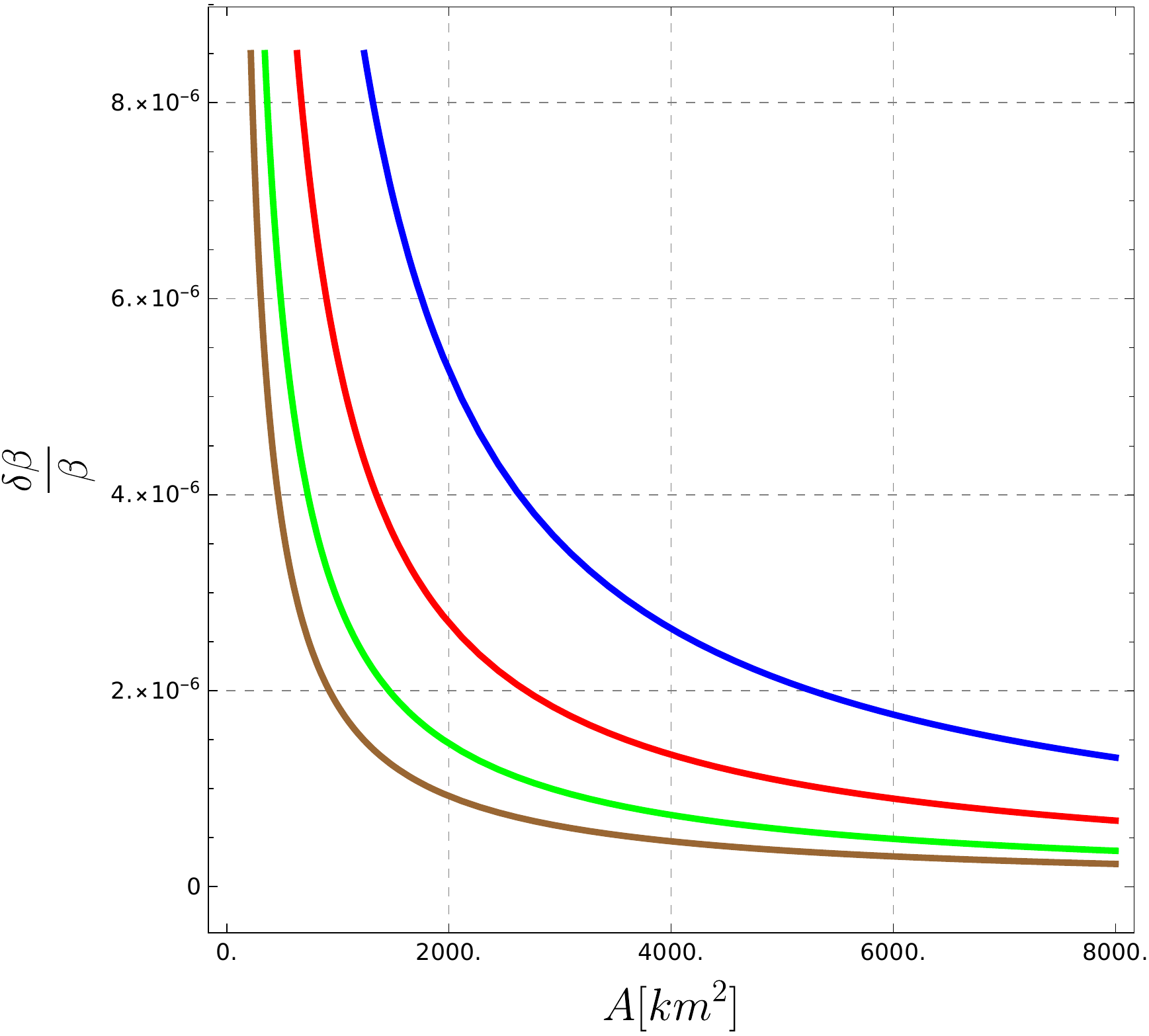} \\
(b)  \\[4pt]
\end{tabular}
\caption{Relative error of the parameters $\gamma$ and $\beta$, as measured by a external clock placed at the detectors, through Eqs.\thinspace(\ref{DeltaTau1}) and \thinspace(\ref{DeltaTau2}) in terms of the area $A = \Delta h \times L_{\gamma_2}$ of the array. (a) Relative error for the parameter $\gamma$, and the clock employed has a uncertainty $\delta \tau_{\rm clock } = 10^{-18} [\rm s]$. (b) Relative error for the parameter $\beta$. We assume that in both HOM array configuration the clocks have the same uncertainty. $\delta \tau_{\rm clock } = 10^{-31} [\rm s]$ . Each array configuration is characterized by an area, $A_1$ is used in the measurement of the first time delay and $A_2 = \eta A_1$ is used in the second array, where $\eta$ is a constant. The blue continuous line represents $\eta =1/8$, red continuous line $\eta=1/4$, green continuous line $\eta=1/2$ and brown continuous line $\eta=1$.}
\label{RelErrorgammabetatauPlot}
\end{figure*}

\begin{figure*}
\centering
\begin{tabular}{c}
\includegraphics[width=90mm]{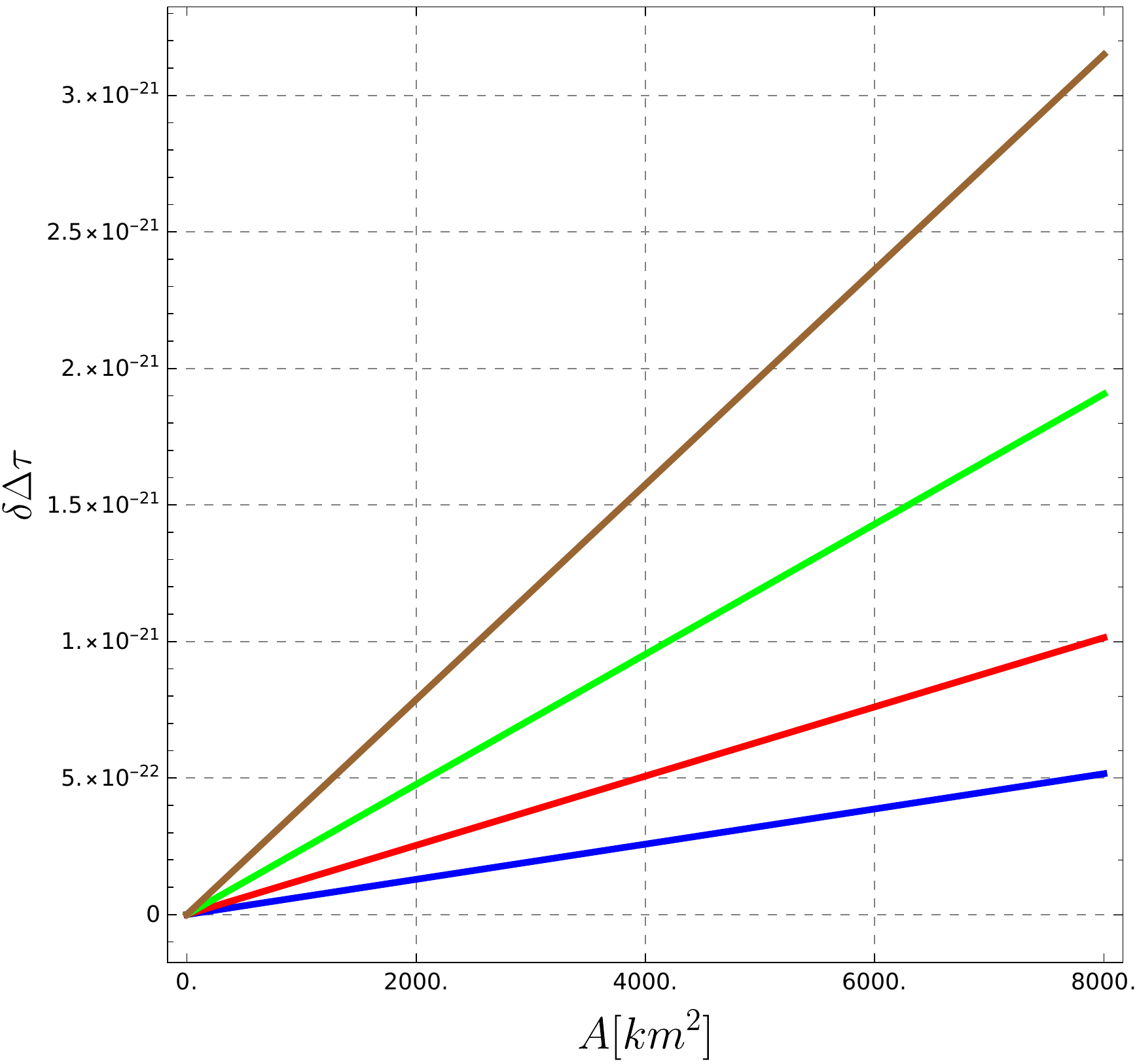} \\
(a) \\
 \includegraphics[width=90mm]{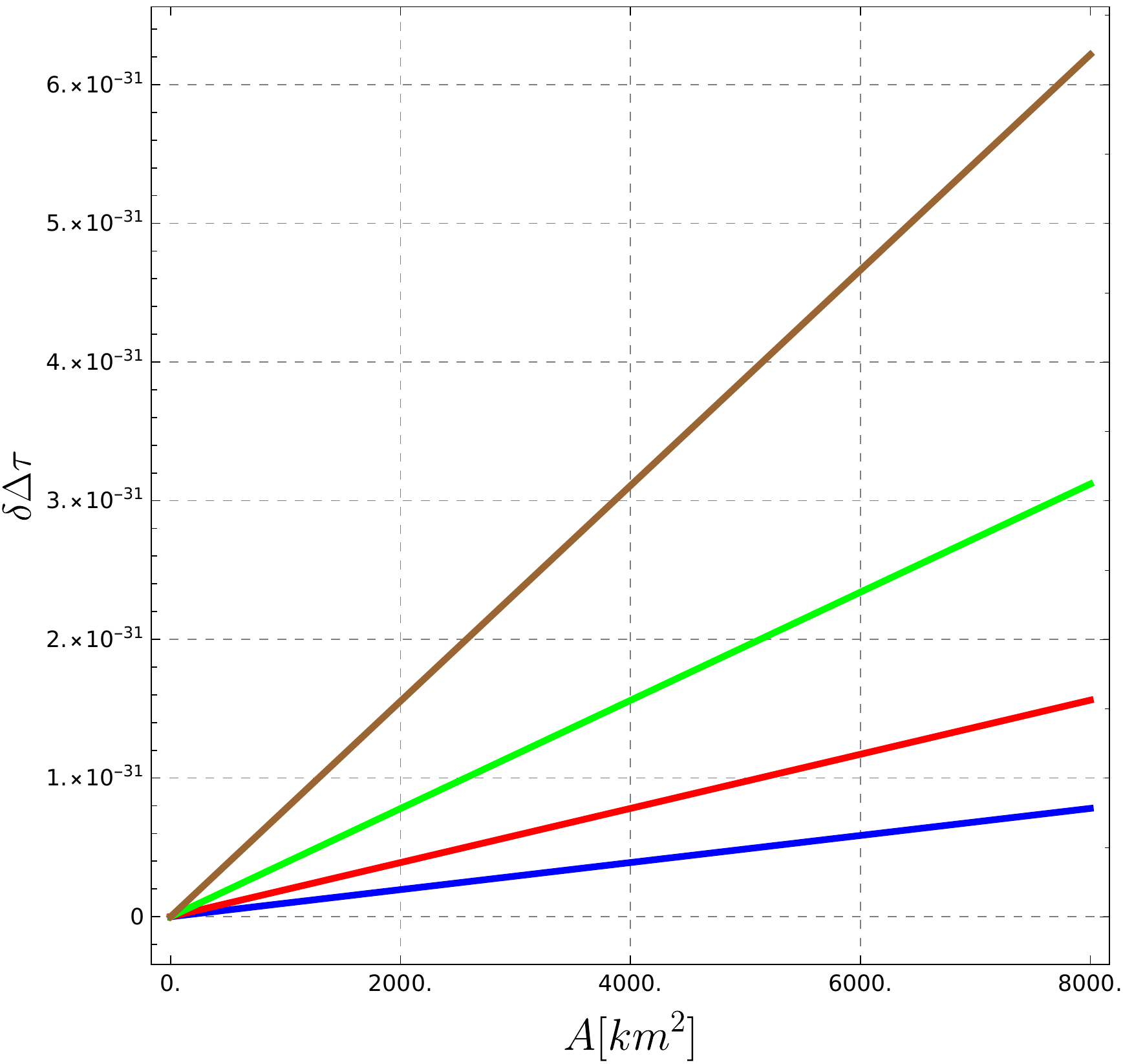} \\
(b)  \\[4pt]
\end{tabular}
\caption{ Uncertainty on the measurement of the temporal delay necessary to obtain the current value in the uncertainty of $\gamma$ and $\beta$, in terms of the area $A = \Delta h \times L_{\gamma_2}$ of the array. (a) Uncertainty of $\Delta \tau$ for the parameter $\gamma$. (b) Uncertainty of $\Delta \tau$ for the parameter $\beta$. We assume that in both HOM array configurations the uncertainty in the arrival time is the same. Each array configuration is characterized by an area, $A_1$ is used for measuring the first time delay and $A_2 = \eta A_1$ is used in the second array, where $\eta$ is a constant. Blue continuous line represents $\eta =1/8$, red continuous line $\eta=1/4$, green continuous line $\eta=1/2$ and brown continuous line $\eta=1$. }
\label{UncertaintyintauvsareaPlotHOM}
\end{figure*}

 To calculate the uncertainty of $\gamma$  and $\beta$ , we consider errors only in the measuring process of the proper time of each HOM interferometer. Moreover the uncertainties of the measurement of the arrival time in each configuration of the HOM array are considered independently, therefore we are assuming that there is no correlation between them. In this case the uncertainty of the parameters $\gamma$ and $\beta$ is given by
\begin{eqnarray}
\delta \epsilon(\Delta \tau_1, \Delta \tau_2) &=& \sqrt{\left(\frac{\partial \epsilon}{\partial \Delta \tau_1} \right)^2 \delta \Delta^2 \tau_1 + \left(\frac{\partial \epsilon}{\partial \Delta \tau_2} \right)^2 \delta \Delta^2 \tau_2},
\end{eqnarray}
where  $\epsilon = (\gamma,\beta)$ corresponds to each parameter $\gamma$ and $\beta$, $\delta \Delta \tau_1$ and $\delta \Delta \tau_2$ are the uncertainties on the measurement of the temporal delays $\Delta \tau_1$ and $\Delta \tau_2$, respectively. Therefore,  given the uncertainties $\delta \Delta \tau_1$ and $\delta \Delta \tau_2$  for the two configurations of the HOM array, the uncertainty in the estimation of $\gamma$ becomes
\begin{eqnarray}
\delta \gamma &=& \frac{c^3}{g L_{\gamma_2} L_{\gamma'_2} \Delta h \left[ \phi(R_1) - \phi(R_2)\right]}\sqrt{\left( L^{2}_{\gamma_2} \delta \Delta \tau^{2}_{1} \phi^{2}(R_1) + L^{2}_{\gamma'_2} \delta \Delta \tau^{2}_{1} \phi^{2}(R_2)\right)}. \nonumber \\
\end{eqnarray}
For the $\beta$ parameter we obtain
\begin{eqnarray}
\delta \beta &=&  \frac{c^3}{2 g^2 L^{2}_{\gamma_2} L^{2}_{\gamma'_2} \Delta h^2 \left[ \phi(R_1) - \phi(R_2) \right]^2 } \nonumber \\
&& \left( L^{2}_{\gamma_2} \delta \Delta^{2} \tau_2 \left[ L_{\gamma_2}\phi(R_1) \left( c^2 g L_{\gamma'_2} \Delta h + 4\phi(R_1)(c^3 \Delta \tau_2 - g L_{\gamma'_2}\Delta h) \right) \right. \right. \nonumber \\
&& \left. \left. - L_{\gamma'_2} \left( c^2 g L_{\gamma_2} \Delta h + 4(c^3 \Delta \tau_1 - g L_{\gamma_2} \Delta h) \phi(R_1)\right) \phi(R_2) \right]^2 \right. \nonumber \\
&& \left.  +  L^{2}_{\gamma'_2} \delta \Delta^{2} \tau_1 \left[ L_{\gamma'_2}\phi(R_2) \left(( c^2 g L_{\gamma_2} \Delta h + 4\phi(R_2)(c^3 \Delta \tau_1 - g L_{\gamma_2}\Delta h) \right) \right) \right. \nonumber \\
&& \left. \left. - L_{\gamma_2} \left( c^2 g L_{\gamma_2} \Delta h + 4(c^3 \Delta \tau_1 - g L_{\gamma_2} \Delta h) \phi(R_2)\right) \phi(R_1) \right]^2 \right)^{1/2}. \nonumber \\
\end{eqnarray}
The relative error for the parameter $\gamma$, in terms of the uncertainties $\delta \Delta \tau_1$ and $\delta \Delta \tau_2$,  reads
\begin{eqnarray}
\frac{\delta \gamma}{\gamma} &=& \frac{\sqrt{c^6 \left( L^{2}_{\gamma_2}\delta^2 \Delta \tau_2 \phi^{2}(R_1) + L'^{2}_{\gamma_2}\delta^2 \Delta \tau_1 \phi^{2}(R_2) \right)}}{L'_{\gamma_2}\left(gL_{\gamma_2} \Delta h -c^3 \Delta \tau_1 \right)\phi(R_2)-L_{\gamma_2}\left(gL'_{\gamma_2} \Delta h -c^3 \Delta \tau_2 \right)\phi(R_1)},
\end{eqnarray}
or in terms of the proper area
\begin{eqnarray}
\frac{\delta \gamma}{\gamma} &=& \frac{\sqrt{c^6 \left( A_1^{2}\delta^2 \Delta \tau_2 \phi^{2}(R_1) + A_2^{2}\delta^2 \Delta \tau_1 \phi^{2}(R_2) \right)}}{A_2\left(gA_1 -c^3 \Delta \tau_1 \right)\phi(R_2)-A_1\left(gA_2 -c^3 \Delta \tau_2 \right)\phi(R_1)}.
\label{RelErrorgammaTau}
\end{eqnarray}
The relative error for the parameter $\beta$ is given by

\begin{eqnarray}
\frac{\delta \beta}{\beta} &=& -\left(c^3 \left(2 \Delta \tau_2 L_{\gamma_2}^2 \phi (R_1)^2 \left(c^3 \Delta \tau_2-2 \Delta h g L'_{\gamma_2}\right) \right. \right. \nonumber \\
&& \left. \left. +L_{\gamma_2} L'_{\gamma_2} \phi (R_1) \left(4 \phi (R_2) \left(c^3 (-\Delta \tau_1) \Delta \tau_2+\Delta h \Delta \tau_2 g L_{\gamma_2}+\Delta h \Delta \tau_1 g L'_{\gamma_2}\right) \right. \right. \right. \nonumber \\
&& \left. \left. \left. +c^2 \Delta h g (\Delta \tau_2 L_{\gamma_2}-\Delta \tau_1 L'_{\gamma_2})\right)+L'_{\gamma_2} \phi (R_2) \left(2 \Delta \tau_1 L'_{\gamma_2} \phi (R_2) \left(c^3 \Delta \tau_1-2 \Delta h g L_{\gamma_2}\right)  \right. \right. \right. \nonumber \\
&& \left. \left. \left. +c^2 \Delta h g L_{\gamma_2} (\Delta \tau_1 L'_{\gamma_2}-\Delta \tau_2 L_{\gamma_2})\right)\right) \right)^{-1} \nonumber \\
&& \times  \left[c^6 \left(\delta \Delta \tau_{2}^2 L^{2}_{\gamma_2} \left(L_{\gamma_2} \phi (R_1) \left(\phi(R_1) \left(4 c^3 \Delta \tau_2 -4 \Delta h g L'_{\gamma_2}\right)+c^2 \Delta h g L'_{\gamma_2}\right) \right. \right. \right. \nonumber \\
&& \left. \left. \left. -L'_{\gamma_2} \phi (R_2) \left(\phi (R_1) \left(4 c^3 \Delta \tau_1-4 \Delta h g L_{\gamma_2}\right)+c^2 \Delta h g L_{\gamma_2}\right)\right)^2 \right. \right. \nonumber \\
&& \left. \left. +\delta \Delta \tau_{1}^2 L^{'2}_{\gamma_2} \left(L'_{\gamma_2} \phi(R_2) \left(\phi (R_2) \left(4 c^3 \Delta \tau_1-4 \Delta h g L_{\gamma_2}\right)+c^2 \Delta h g L_{\gamma_2}\right) \right. \right. \right. \nonumber \\
&& \left. \left. \left. -L_{\gamma_2} \phi (R_1) \left(\phi (R_2) \left(4 c^3 \Delta \tau_2-4 \Delta h g L'_{\gamma_2}\right)+c^2 \Delta h g L'_{\gamma_2}\right)\right)^2\right)\right]^{1/2}.
\end{eqnarray}
This expression can be written in terms of the proper area $A$ of each array as
\begin{eqnarray}
\frac{\delta \beta}{\beta} &=& -\left(c^3 \left(2 \Delta \tau_2 A_{1}^2 \phi (R_1)^2 \left(c^3 \Delta \tau_2-2  g A_2\right) \right. \right. \nonumber \\
&& \left. \left. + A_1 A_2 \phi (R_1) \left(4 \phi (R_2) \left(c^3 (-\Delta \tau_1) \Delta \tau_2+ \Delta \tau_2 g A_1 + \Delta \tau_1 g A_2\right) +c^2  g (\Delta \tau_2 A_1 -\Delta \tau_1 A_2)\right) \right. \right. \nonumber \\
&& \left. \left. +A_2 \phi (R_2) \left(2 \Delta \tau_1 A_2 \phi (R_2) \left(c^3 \Delta \tau_1-2  g L_{\gamma_2}\right)
 +c^2 \Delta h g A_1 (\Delta \tau_1 A_2 -\Delta \tau_2 A_1)\right)\right) \right)^{-1} \nonumber \\
&& \times  \left[c^6 \left(\delta \Delta \tau_{2}^2 A_1^{2} \left(A_1 \phi (R_1) \left(\phi(R_1) \left(4 c^3 \Delta \tau_2 -4  g A_2 \right)+c^2  g A_2 \right) \right. \right. \right. \nonumber \\
&& \left. \left. \left. -A_2 \phi (R_2) \left(\phi (R_1) \left(4 c^3 \Delta \tau_1-4  g A_1\right)+c^2  g A_2 \right)\right)^2 \right. \right. \nonumber \\
&& \left. \left. +\delta \Delta \tau_{1}^2 A_2^{2} \left(A_2 \phi(R_2) \left(\phi (R_2) \left(4 c^3 \Delta \tau_1-4 g A_1 \right)+ c^2  g A_1 \right) \right. \right. \right. \nonumber \\
&& \left. \left. \left. -A_1 \phi (R_1) \left(\phi (R_2) \left(4 c^3 \Delta \tau_2-4  g A_2 \right)+c^2  g A_2 \right)\right)^2\right)\right]^{1/2}.
\label{RelErrorbetaTau}
\end{eqnarray}	
 Fig.\thinspace\ref{RelErrorgammabetatauPlot} shows the relative error of the parameters $\gamma$ and $\beta$ in accordance with Eqs.\thinspace(\ref{RelErrorgammaTau}) and (\ref{RelErrorbetaTau}), respectively, in terms of the proper area of the first configuration. In the simulation we consider an external clock that measures the arrival times of the photons,  which has an uncertainty $\delta \tau_{\rm clock} = 10^{-18}\;[s]$, and we consider the same uncertainty for both configurations of the HOM array. In order to simplify the analysis we configure the area of the second array as $A_2 = \eta \times A_1$, with $\eta=1,1/2,1/4,1/8$. As we can see from the figure, the configuration with a lower relative error for the parameters $\gamma$ and $\beta$ is reached with $A_1 = A_2$. In Fig.~\ref{UncertaintyintauvsareaPlotHOM} the uncertainty  $\delta \Delta\tau$ necessary to obtain a relative error of the parameters $\gamma$ and $\beta$ of $10^{-5}$ is shown. In this case, the configuration of the HOM array with the higher $\delta \Delta \tau$ corresponds to both interferometers having the same area. From the figure, the uncertainty $\delta \Delta \tau$ for $\gamma$ is $\sim 10^{-22}$, while for $\beta$ it corresponds to $\sim 10^{-31}$. In this case, to obtain the current uncertainties of $\gamma$ and $\beta$ we need a lower error in the arrival times for $\beta$.

\subsection{Errors in the coincidence detection probability}

\begin{figure*}
\centering
\begin{tabular}{c}
\includegraphics[width=90mm]{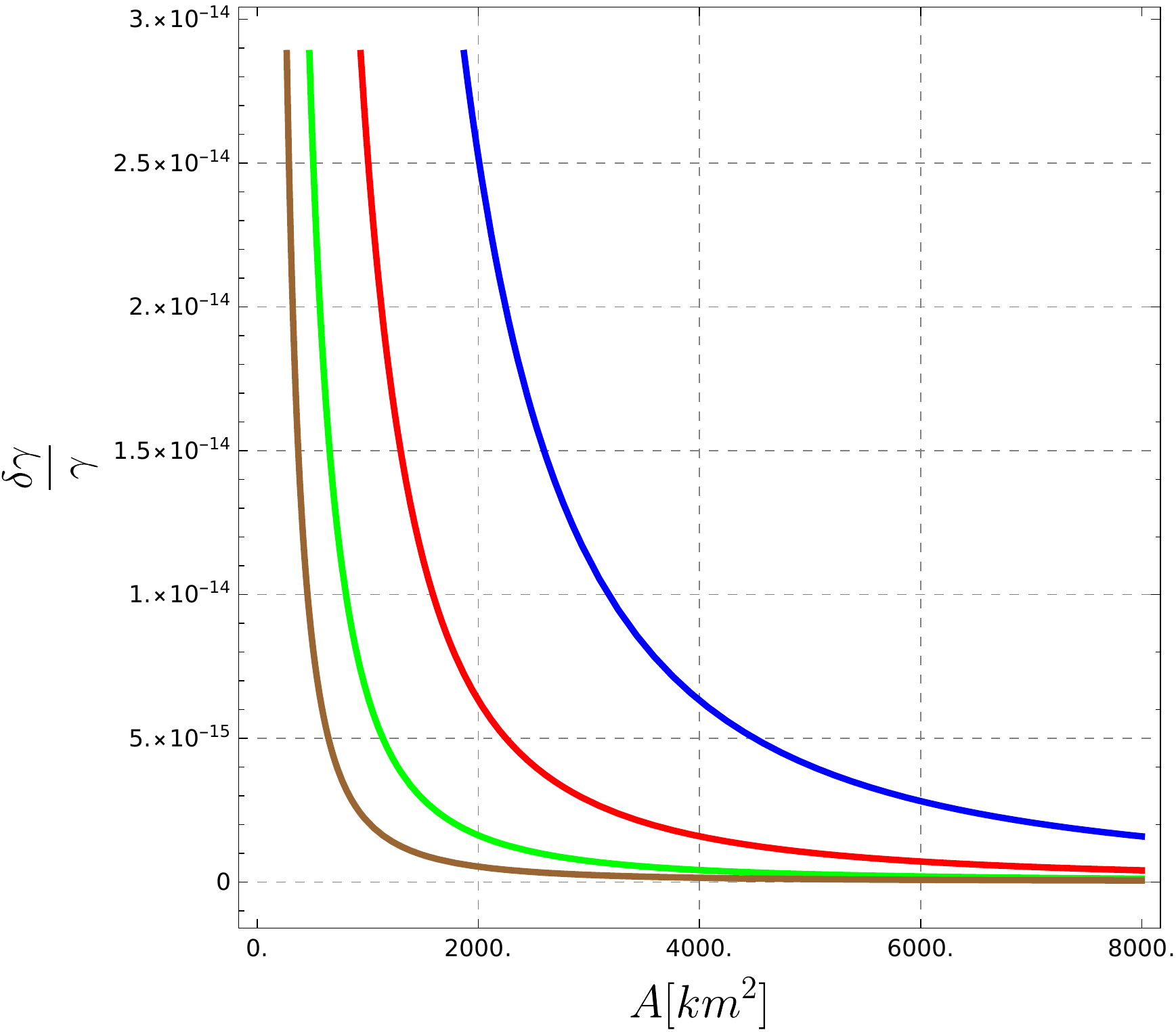} \\
(a) \\
 \includegraphics[width=90mm]{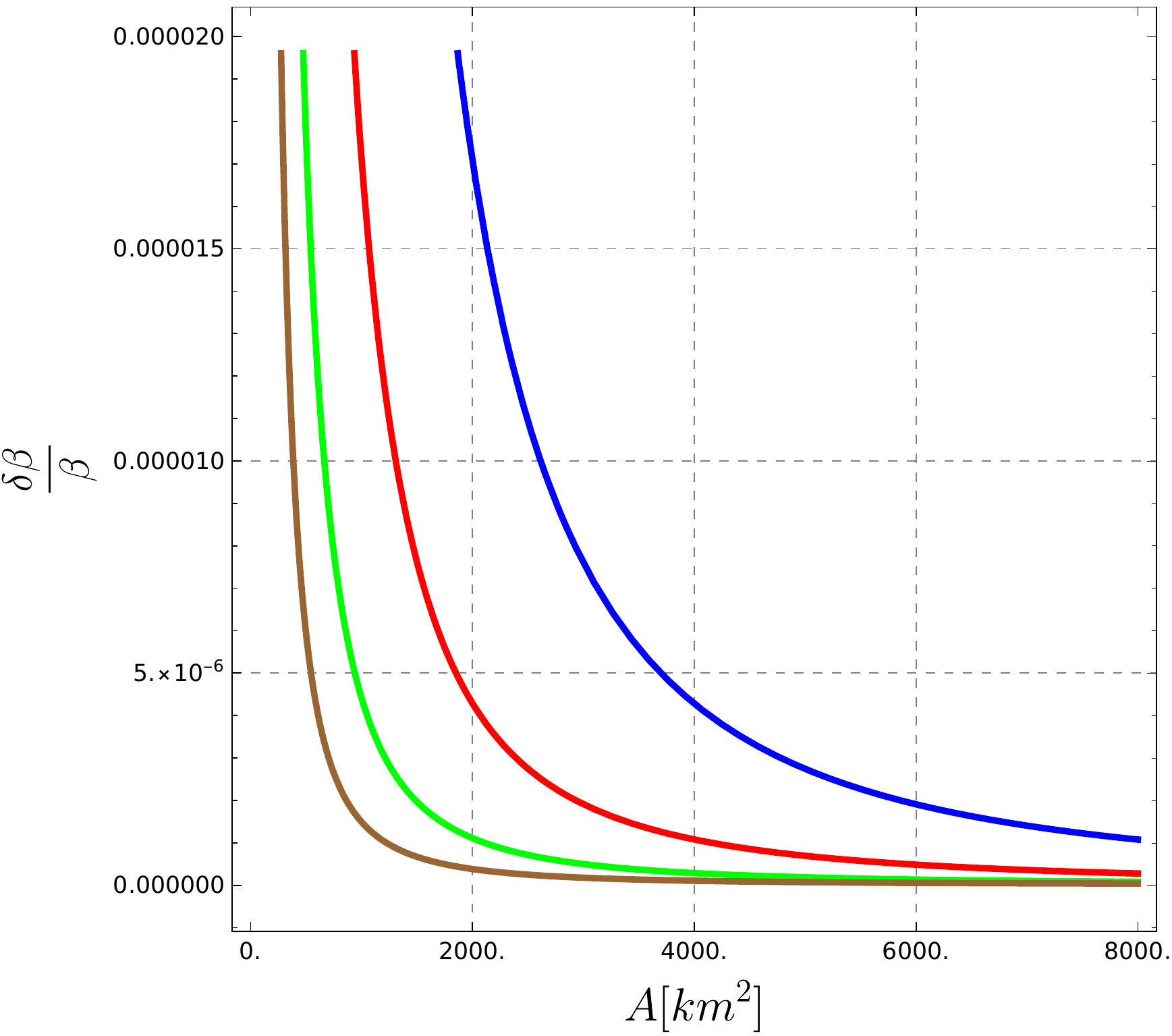} \\
(b)  \\[4pt]
\end{tabular}
\caption{Relative error of the parameters $\gamma$ and $\beta$ measured through the probability of detection Eq.\thinspace(\ref{ProbCoinc}), in terms of the area $A = \Delta h \times L_{\gamma_2}$ of the array. (a) Relative error for the parameter $\gamma$. (b) Relative error for the parameter $\beta$. For both simulations the uncertainty in the measure of the probability of detection is $\delta S = 10^{-18}$. We assume that in both HOM array configuration the uncertainty in the probability of detection is the same. In the array configuration, $A_1$ is the area for the first array configuration, in which the detection of the first time delay is done. The second array has area $A_2 = \eta A_1$, where $\eta$ is a constant. Blue continuous line represents $\eta =1/8$, red continuous line $\eta=1/4$, green continuous line $\eta=1/2$ and brown continuous line $\eta=1$. }
\label{RelErrorgammabetatauPlotHOM}
\end{figure*}

\begin{figure*}
\centering
\begin{tabular}{c}
\includegraphics[width=90mm]{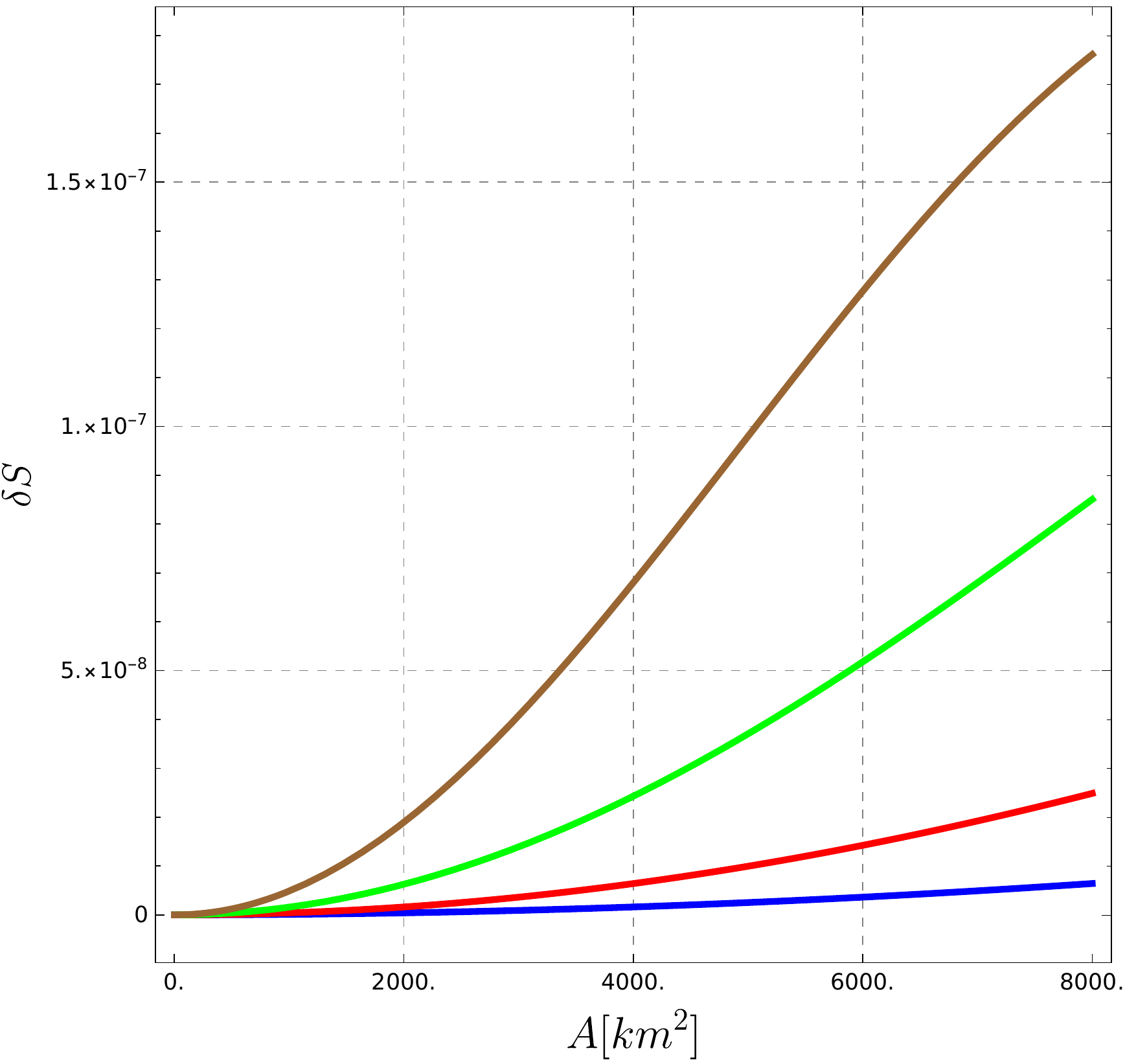} \\
(a) \\
 \includegraphics[width=90mm]{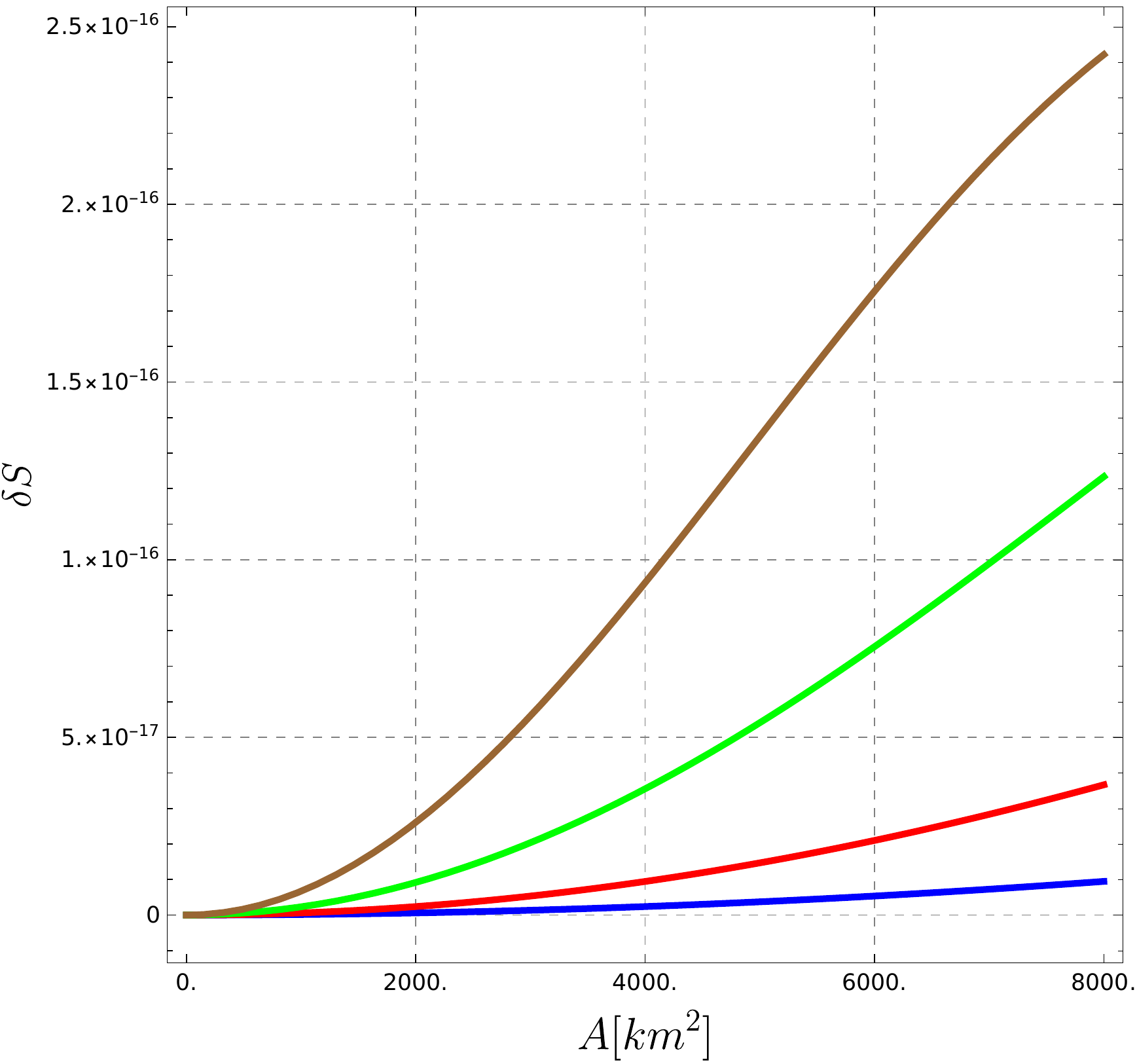} \\
(b)  \\[4pt]
\end{tabular}
\caption{ Uncertainty on the measurement of the coincidence observable S Eq.\thinspace(\ref{CoincOperator}) necessary to obtain the current value in the uncertainty of $\gamma$ and $\beta$, in terms of the area $A = \Delta h \times L_{\gamma_2}$ of the array. (a) Uncertainty of S for the parameter $\gamma$. (b) Uncertainty of S for the parameter $\beta$. We assume that in both HOM array configuration the uncertainty in the probability of detection is the same. In the array configuration, $A_1$ is the area for the first array configuration, in which the detection of the first time delay is done. The second array has area $A_2 = \eta A_1$, where $\eta$ is a constan. Blue continuous line represents $\eta =1/8$, red continuous line $\eta=1/4$, green continuous line $\eta=1/2$ and brown continuous line $\eta=1$. }
\label{RelErrorgammabetataudeltaS}
\end{figure*}

Considering the expectation value of the observable of detection in coincidence for two photons Eq.\thinspace(\ref{ProbObs}), the uncertainty for the parameter $\gamma$ reads
\begin{eqnarray}
\delta \gamma &=& \sqrt{\frac{c^6 \left(\frac{\delta S_1^2 \phi (R_2)^2}{A_1^2 (1 - 2 \langle S_1 \rangle)^2 \sigma_1^2 \ln \left(\frac{1}{1-2 \langle S_1 \rangle}\right)}+\frac{\delta S_2^2 \phi (R_1)^2}{A_2^2 (1 - 2 \langle S_2 \rangle)^2 \sigma_2^2 \ln \left(\frac{1}{1-2 \langle S_2 \rangle}\right)}\right)}{g^2 (\Delta \phi_{12})^2}},
\label{ErrorgammaProbHOM}
\end{eqnarray}
and for the parameter $\beta$ 
\begin{eqnarray}
\delta \beta &=& \frac{1}{2 A_1^2 A_2^2 g^2 \sigma_1^2 \sigma_2^2 (\Delta \phi_{12})^2} \nonumber \\
&& \times \left(c^6 \left(\frac{A_1^2 \delta S_2^2 \sigma_1^2}{(1-2 \langle S_2 \rangle)^2 \ln \left(\frac{1}{1-2 \langle S_2 \rangle}\right)} \left(A_1 \sigma_1 \phi (R_1) \left(\phi (R_1) \left(4 c^3 \sqrt{\ln \left(\frac{1}{1-2 \langle S_2 \rangle}\right)}-4 A_2 g \sigma_2\right)\right. \right. \right. \right. \nonumber \\
&& \left. \left. \left. \left.  +A_2 c^2 g \sigma_2\right)-A_2 \sigma_2 \phi (R_2) \left(\phi (R_1) \left(4 c^3 \sqrt{\ln \left(\frac{1}{1-2 \langle S_1 \rangle}\right)}-4 A_1 g \sigma_1\right)+A_1 c^2 g \sigma_1\right)\right)^2 \right. \right. \nonumber \\
&& \left. \left. +\frac{A_2^2 \delta S_1 ^2 \sigma_2^2}{(1-2 \langle S_1 \rangle)^2 \ln \left(\frac{1}{1-2 \langle S_1 \rangle}\right)} \left(A_2 \sigma_2 \phi (R_2) \left(4 \phi (R_2) \left(A_1 g \sigma_1-c^3 \sqrt{\ln \left(\frac{1}{1-2 \langle S_1 \rangle}\right)}\right)-A_1 c^2 g \sigma_1\right) \right. \right. \right. \nonumber \\
&& \left. \left. \left.  +A_1 \sigma_1 \phi (R_1) \left(\phi (R_2) \left(4 c^3 \sqrt{\ln \left(\frac{1}{1-2 \langle S_2 \rangle}\right)}-4 A_2 g \sigma_2\right)+A_2 c^2 g \sigma_2\right)\right)^2\right)\right)^{1/2}.
\label{ErrorbetaProbHOM}
\end{eqnarray}
Considering the errors Eqs.~(\ref{ErrorgammaProbHOM}) and (\ref{ErrorbetaProbHOM}) together with Eqs.~(\ref{ParamterGamma2}) and (\ref{ParamterBeta}) we can calculate the relative error of the parameters $\gamma$ and $\beta$, we do not show the expression because it is not enlightening. In Fig.~\ref{RelErrorgammabetatauPlotHOM} we show the relative error of the parameters $\gamma$ and $\beta$. We choose the same configuration of the HOM arrays as in the previous section, as indicated in the figure. The lower relative error for both parameters is reached when both areas are equal. In our simulations we employ states with wavelength $\lambda = 0.995\;[\mu m]$ and bandwidth $\delta \lambda = 0.034 [\mu m]$ employed in current SPDC sources \cite{Vanselow-2019-SPDC}, and an error in the measurement of the coincidence operator $\delta S = 10^{-18}$. Therefore, the relative error of $\gamma$ and $\beta$ are $\sim 10^{-14}$ and $\sim 10^{-6}$, respectively.  Fig. \thinspace\ref{RelErrorgammabetataudeltaS} shows the uncertainty $\delta S$ of the measurement of the operator detection in coincidence if we consider the errors currently indicated in the literature, in terms of the effective area of one of the arrays. As in the previous section, considering both configurations of the interferometer with the same proper area we obtain a higher uncertainty $\delta S$. For $\gamma$ the uncertainty $\delta S \sim 10^{-7}$, and for $\beta$ $\delta S \sim 10^{-16}$.  


\subsection{Relative error employing a two-mode squeezed-vacuum state}

\begin{figure*}
\centering
\begin{tabular}{c}
\includegraphics[width=90mm]{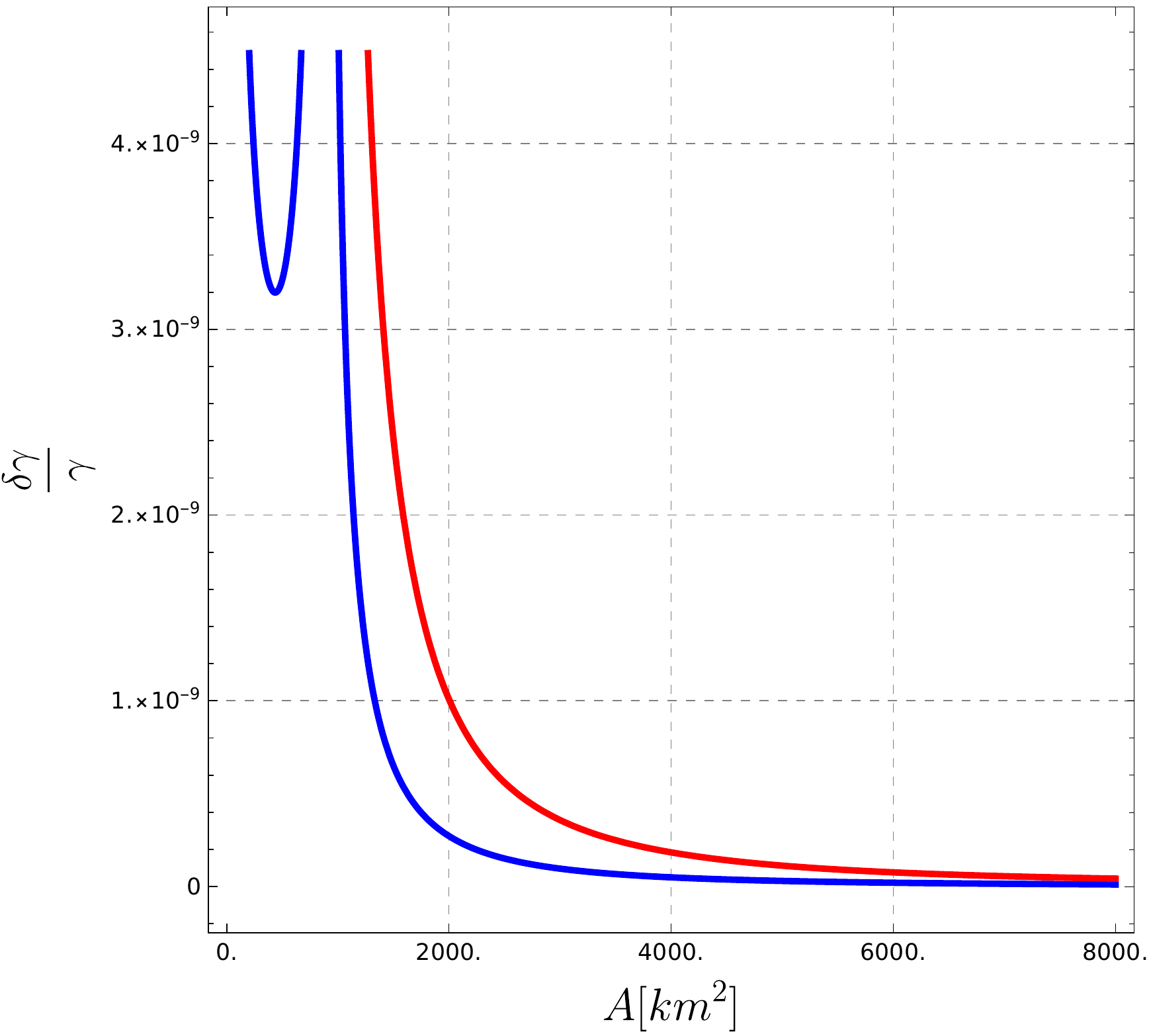} \\
(a) \\
 \includegraphics[width=90mm]{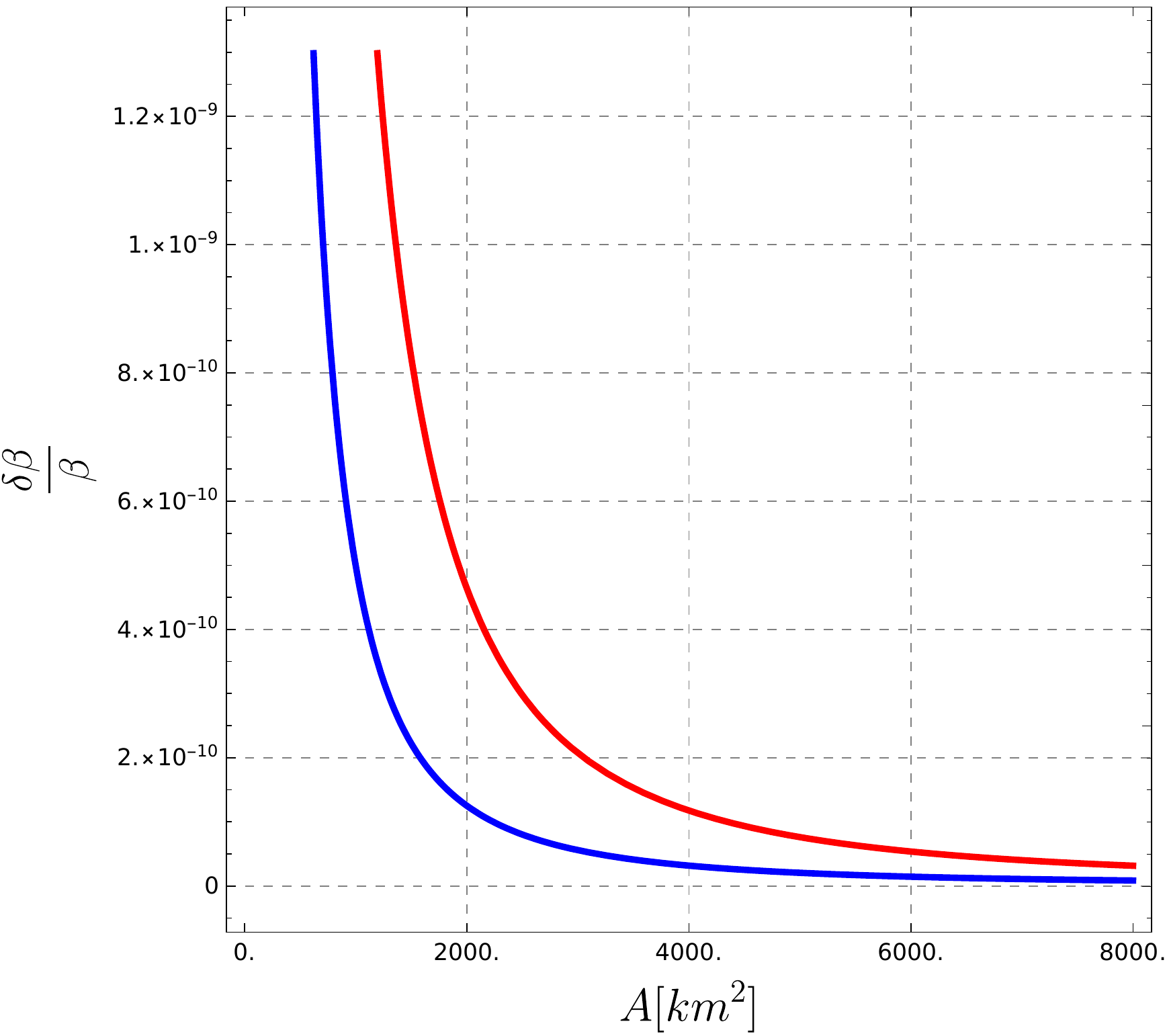} \\
(b)  \\[4pt]
\end{tabular}
\caption{Relative error of the parameters $\gamma$ and $\beta$ measured through the probability of detection Eq.\thinspace(\ref{SingleVacuumStateProbability}), in terms of the area $A = \Delta h \times L_{\gamma_2}$ of the array. (a) Relative error parameter $\gamma$. (b) Relative error parameter $\beta$. For both simulations the uncertainty in the measure of the probability of detection is $\delta S = 10^{-13}$. We assume that in both HOM array configuration the uncertainty in the probability of detection is the same. In the array configuration, given an area $A_1$ for the detection of the first time delay, the second area $A_2 = \eta A_1$, where $\eta=1/4$. Blue continuous line represents the squeezing parameter $r=1$, red continuous line $r=2$. The mean wavelength of both photons is $\lambda = 0.995 \;[\mu m]$ and a bandwidth $\delta \lambda = 0.034\;[\mu m]$. }
\label{RelErrorgammabetatauPlotSqueezing}
\end{figure*}

\begin{figure*}
\centering
\begin{tabular}{c}
\includegraphics[width=90mm]{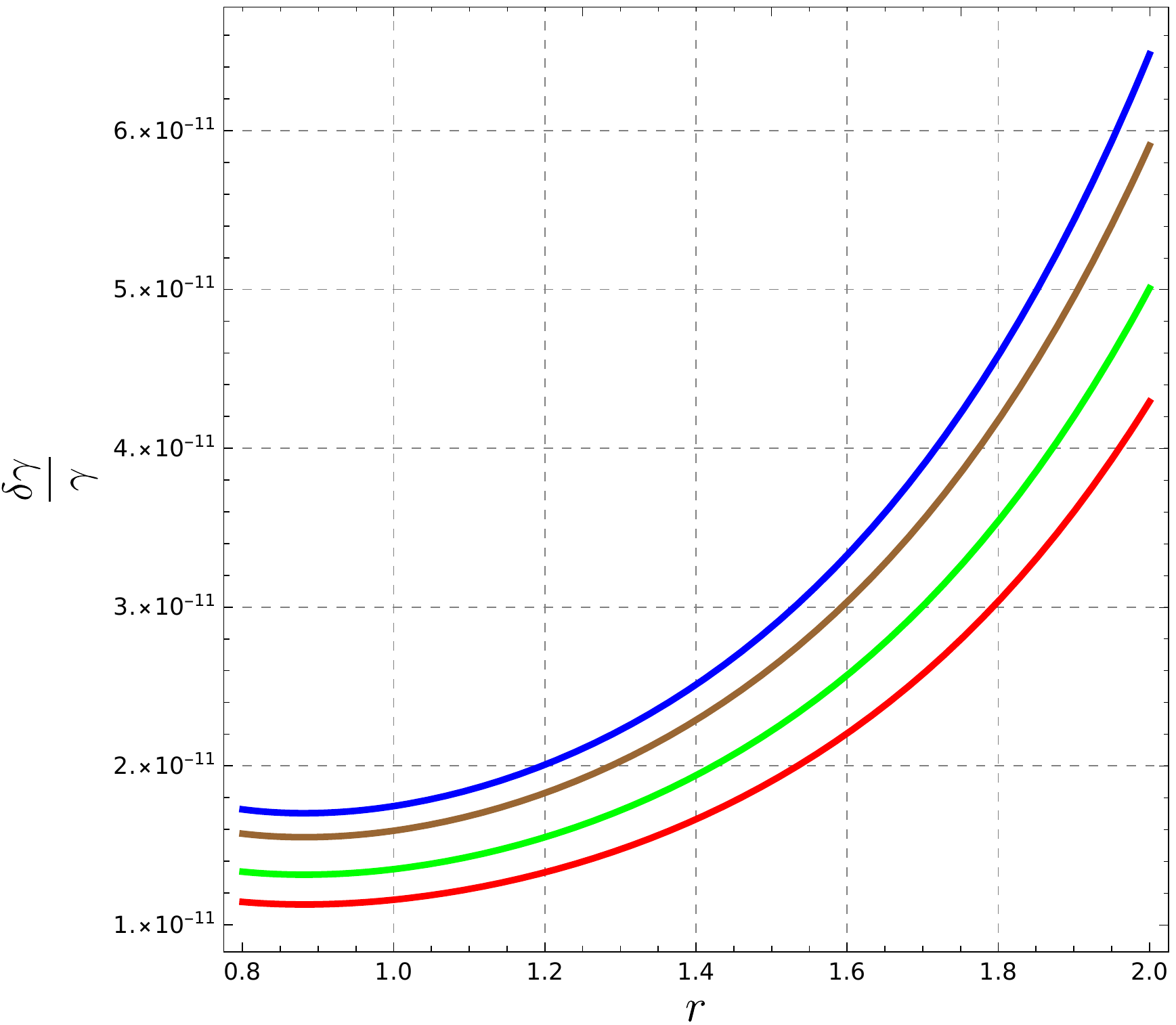} \\
(a) \\
 \includegraphics[width=90mm]{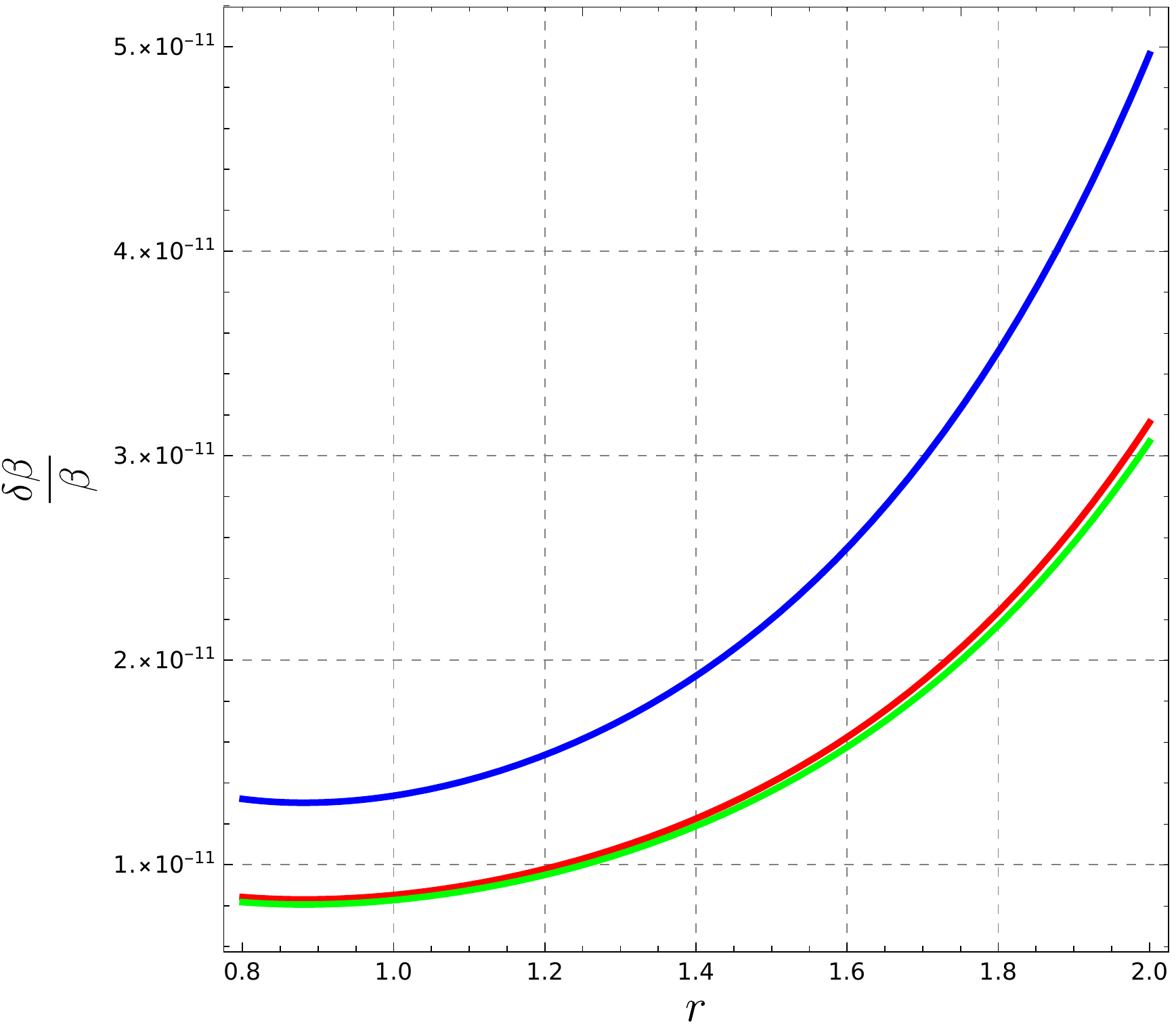} \\
(b)  \\[4pt]
\end{tabular}
\caption{Relative error of the parameter $\gamma$ and $\beta$ measured through the probability of detection Eq.\thinspace(\ref{SingleVacuumStateProbability}), in terms of the squeezing parameter $r$. (a) Relative error for the parameter $\gamma$. (b) Relative error for the parameter $\beta$. For both simulations the uncertainty in the measure of the probability of detection is $\delta S = 10^{-13}$. We assume that, in both HOM array configurations, the uncertainty in the probability of detection is the same. In the array configuration, given an area $A_1$ for the detection of the first time delay, the second area $A_2 = \eta A_1$, where $\eta$ is a constant.  Continuous blue line $\eta =1/8$, continuous red line $\eta=1/4$, continuous green line $\eta=1/2$. For (a) Continuous brown line $\eta=1$.  The mean wavelength of both photons is $\lambda = 0.995 \;[\mu m]$ and a bandwidth $\delta \lambda = 0.034\;[\mu m]$. }
\label{RelErrorgammabetaSqueezing}
\end{figure*}

Here, we calculate the relative error of the parameters $\gamma$ and $\beta$ given in Eqs.\thinspace(\ref{gammaSqueezing}) and (\ref{betaSqueezing}), respectively.  The uncertainty of the parameter $\gamma$ is given by
\begin{eqnarray}
\delta \gamma &=&\frac{\sqrt{2} c^3}{A_1 A_2 g \sigma (\phi (R_1)-\phi (R_2))} \nonumber \\
&& \times \left(A_1 \phi(R_1) \ln \left(1-2 \langle S_2 \rangle \left(\tanh^2(r_2)+1\right)^2\right)-A_2 \phi (R_2) \ln \left(1-2 \langle S_1 \rangle \left(\tanh ^2(r_1)+1\right)^2\right)\right)-1 \nonumber \\
\label{Errorgammasqz}
\end{eqnarray}
where $\delta S_1$ and $\delta S_2$ are the uncertainties on the measurements of the probabilities $\langle S_1 \rangle$ and $\langle S_2 \rangle$, respectively.  $r_1$ and $r_2$ are the squeezing parameters employed in the generation of each squeezed single-vacuum state for each configuration of the HOM array, respectively. For the parameter $\beta$ we obtain
\begin{eqnarray}
\delta \beta &=& \frac{c^3}{2 A_1^2 A_2^2 g^2 \sigma ^2 (\phi (R_1)-\phi (R_2))^2} \nonumber \\
&& \left(-4 c^3 \left(A_1 \phi (R_1) \ln \left(1-2 \langle S_2 \rangle \left(\tanh ^2(r_2)+1\right)^2\right)-A_2 \phi (R_2) \ln \left(1-2 \langle S_1 \rangle \left(\tanh ^2(r_1)+1\right)^2\right)\right)^2 \right. \nonumber \\
&& \left. +\sqrt{2} A_1 A_2 c^2 g \sigma  (\phi (R_1)-\phi (R_2)) \left(A_2 \ln \left(1-2 \langle S_1 \rangle \left(\tanh ^2(r_1)+1\right)^2\right) \right. \right. \nonumber \\
&& \left. \left. -A_1 \ln \left(1-2 \langle S_2 \rangle \left(\tanh ^2(r_2)+1\right)^2\right)\right) \right. \nonumber \\
&& \left. +4 \sqrt{2} A_1 A_2 g \sigma  (\phi (R_1)-\phi (R_2)) \left(A_1 \phi (R_1) \ln \left(1-2 \langle S_2 \rangle \left(\tanh ^2(r_2)+1\right)^2\right) \right. \right. \nonumber \\
&& \left. \left. -  \phi (R_2) \ln \left(1-2 \langle S_1 \rangle \left(\tanh ^2(r_1)+1\right)^2\right)\right)\right).
\label{Errorbetasqz}
\end{eqnarray}
The relative error of the parameter $\gamma$ is obtained through the coefficient between Eqs.\thinspace(\ref{Errorgammasqz}) and (\ref{gammaSqueezing}). Analogously, $\beta$ is given by comparing Eqs.\thinspace(\ref{Errorbetasqz}) and (\ref{betaSqueezing}).  Fig.\thinspace\ref{RelErrorgammabetatauPlotSqueezing} shows the relative error for the parameters $\gamma$ and $\beta$ in terms of the effective area of the HOM array, considering an interval of the squeezing parameter $r$ in the interval $\left[1,2 \right]$ and with an uncertainty $\delta S = 10^{-13}$. Moreover, we consider that in both configurations of the array, the squeezing parameters are equal to $r$. In this sense,  we obtain a lower relative error for both parameters when $r=1$, this is, increasing the squeezing parameter does not reduce relative errors. In this case we have employed the same analysis as before i.e. the area of the second configuration of the array is given by $A_2 = \eta A_1$. Our simulation shows that the lower relative error is found when using $\eta=1/4$ i.e. the area of the second configuration of the HOM array is a quarter of the first area of the array. As in the previous section, we have employed a mean wavelength $\lambda = 0.995\;[\mu m]$ and bandwidth $\delta \lambda = 0.034\;[\mu m]$. In Fig.\thinspace\ref{RelErrorgammabetaSqueezing} we show the relative error of the parameters $\gamma$ and $\beta$ in terms of the squeezing parameter $r$. In this case, we consider a fixed area $A_{\rm max}=8*10^3\;[km^2]$, the same wavelength $\lambda$, and bandwidth $\delta \lambda$ as before, therefore the optimal configuration corresponds to $r \in \left[ 0.8,1.0\right]$, for $\eta= 1/4$ and $1/2$. \\
We now consider two squeezing parameters, $r_1$ and $r_2$, belonging to the first and second array configuration, respectively. Furthermore, we assume a fixed area $A_{\rm max}$ and the same SPDC source as before. In this case, the optimal values  for $r_1$ and $r_2$ belong to the interval $\left[ 0.5,1.0 \right]$. With this configuration, $\gamma$ and $\beta$ achieve relative errors $\sim 10^{-11}$.\\

\begin{figure*}
\centering
\begin{tabular}{c}
\includegraphics[width=90mm]{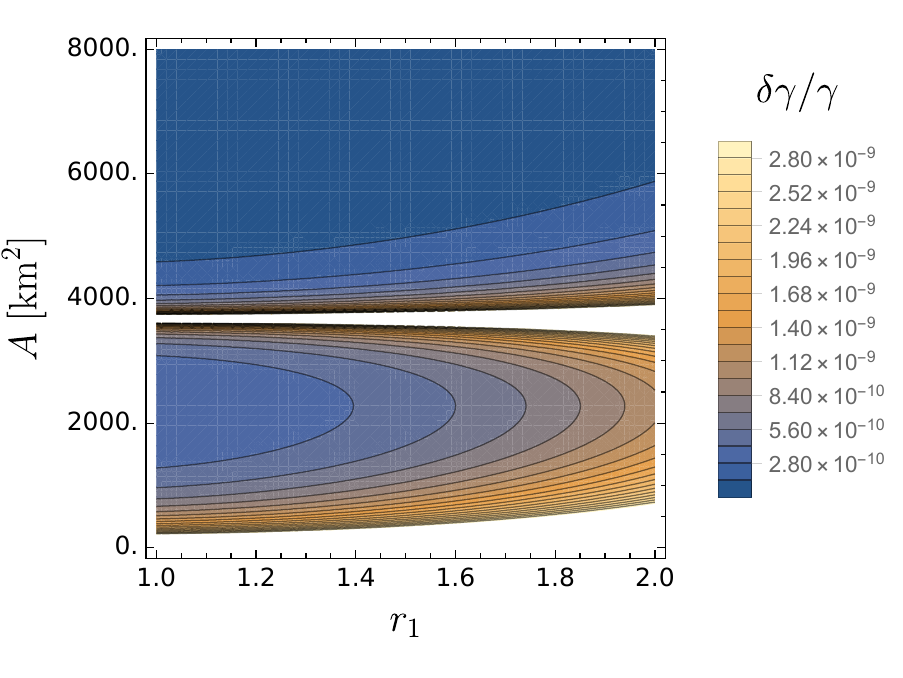} \\
(a) \\
 \includegraphics[width=90mm]{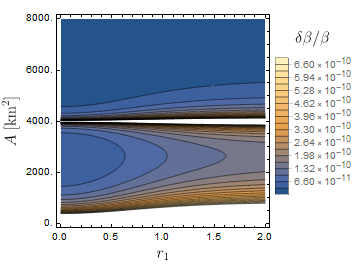} \\
(b)  \\[4pt]
\end{tabular}
\caption{Contour plot of the relative error of the parameter $\gamma$ measured through the probability of detection  Eq.\thinspace(\ref{SingleVacuumStateProbability}), in terms of the squeezing parameters $r_1$ ( assumed to be equal in both configurations of the arrays), and the area $A =\Delta h \times L_{\gamma_2}$.  The uncertainty in the measure of the probability of detection is $\delta S = 10^{-13}$. We assume that, in both HOM array configurations, the uncertainty in the probability of detection is the same.In the array configuration, $A_1$ is the area for the first array configuration, in which the detection of the first time delay is done. The second array has area $A_2 = \eta A_1$, with $\eta=1/4$. The mean wavelength of both photons is $\lambda = 0.995 \;[\mu m]$ and a bandwidth $\delta \lambda = 0.034\;[\mu m]$. }
\label{CPRelErrorgammabetatauPlotSqueezingr1Area}
\end{figure*}

\begin{figure*}
\centering
\begin{tabular}{c}
\includegraphics[width=90mm]{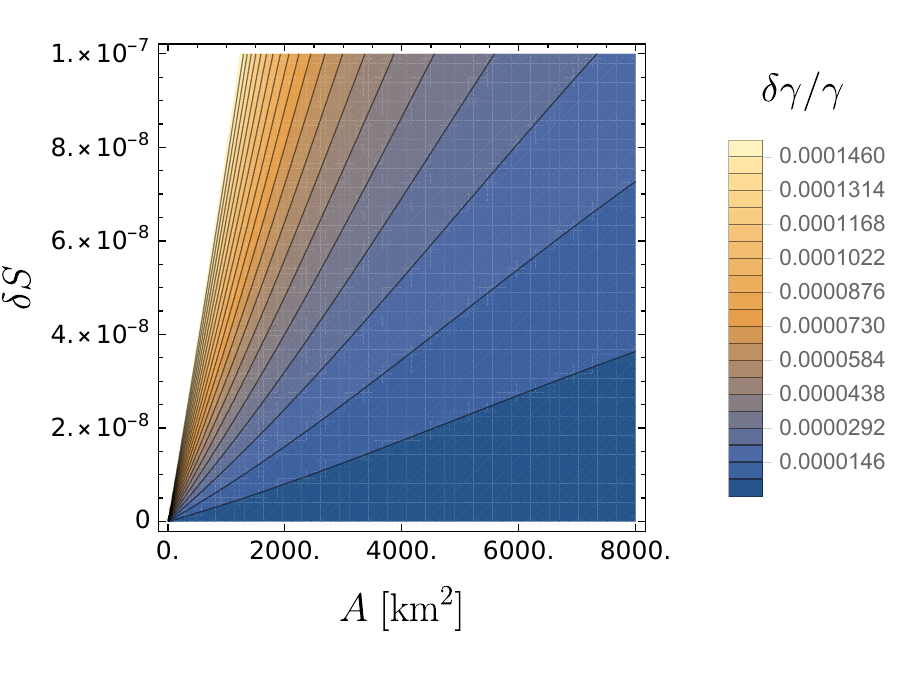} \\
(a) \\
 \includegraphics[width=90mm]{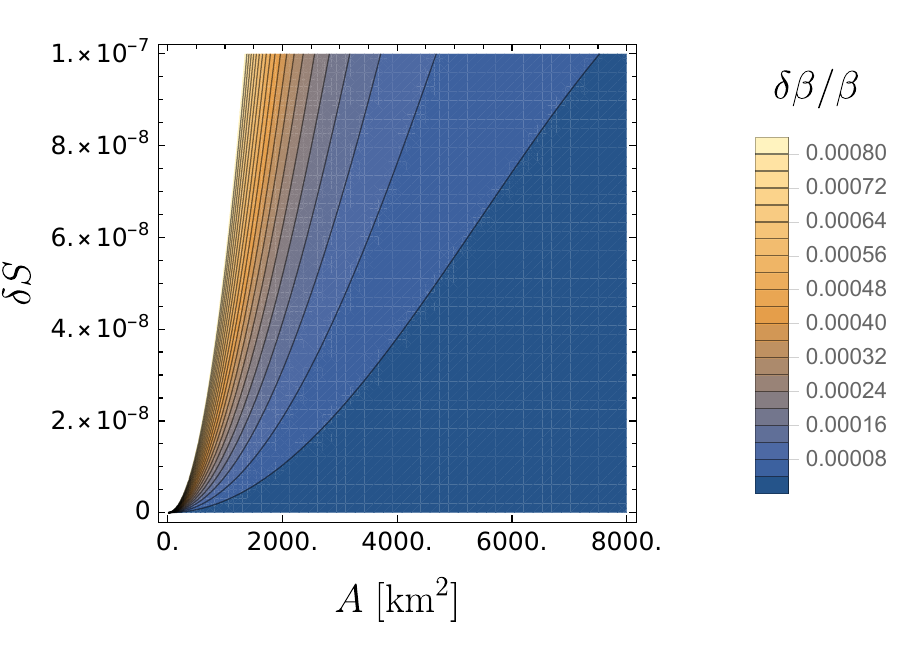} \\
(b)  \\[4pt]
\end{tabular}
\caption{Contour plot of the relative error of the parameter $\gamma$ and $\beta$  measured through the probability of detection  Eq.\thinspace(\ref{SingleVacuumStateProbability}), for a squeezing parameter $r=1$ (assumed to be equal in both configurations of the arrays), in terms of the area $A =\Delta h \times L_{\gamma_2}$ and the uncertainty of the measurement of the coincidence operator $\delta S$ Eq.\thinspace(\ref{CoincOperator}). We assume that in both HOM array configurations, the uncertainty in the probability of detection is the same.In the array configuration, $A_1$ is the area for the first array configuration, in which the detection of the first time delay is done. The second array has area $A_2 = \eta A_1$, with $\eta=1/4$. The mean wavelength of both photons is $\lambda = 0.995 \;[\mu m]$ and a bandwidth $\delta \lambda = 0.034\;[\mu m]$. }
\label{CPRelErrorgammabetaSquezingPlot1AreaDeltaS}
\end{figure*}

\begin{figure*}
\centering
\begin{tabular}{c}
\includegraphics[width=90mm]{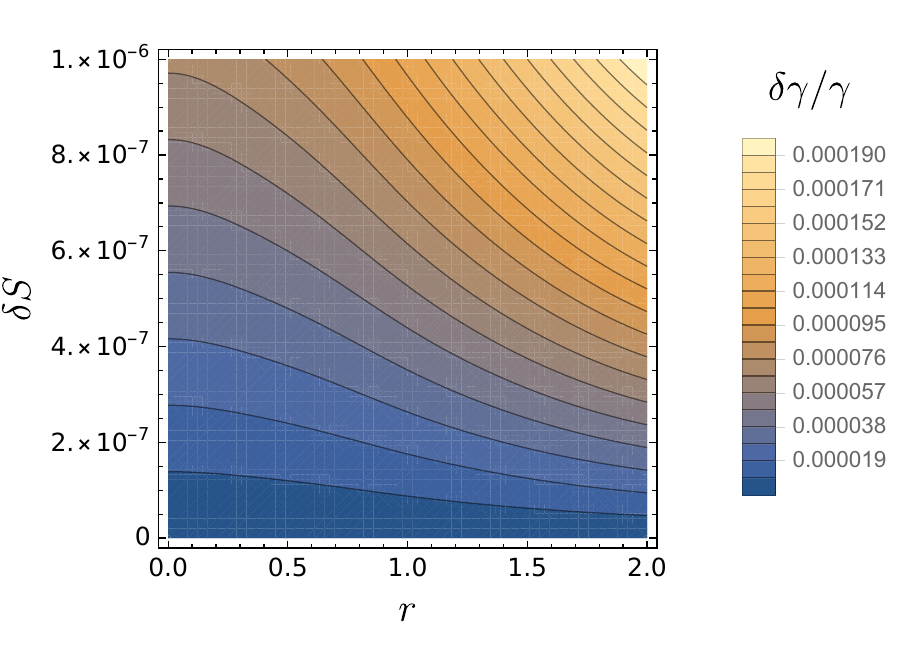} \\
(a) \\
 \includegraphics[width=90mm]{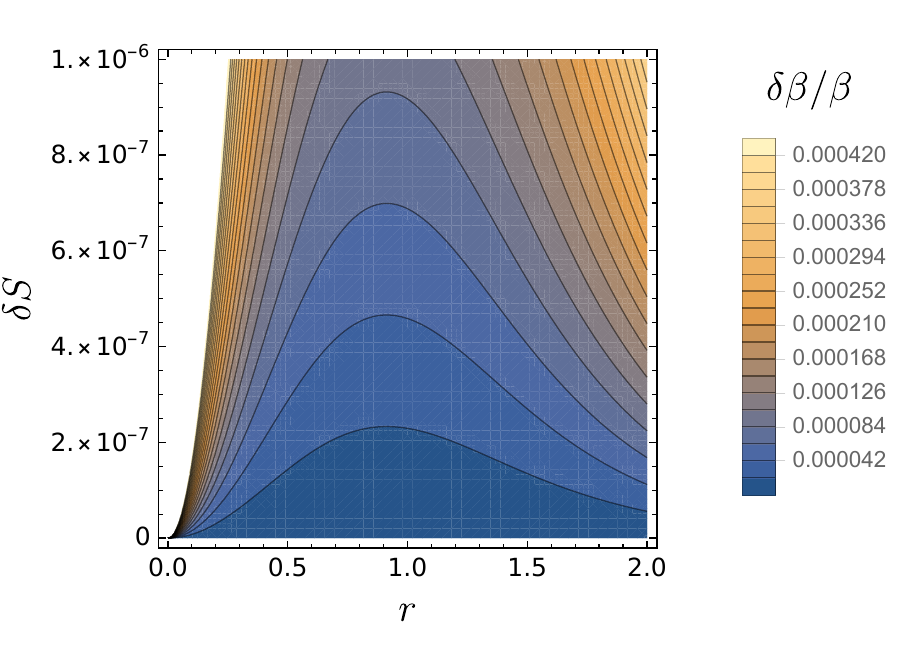} \\
(b)  \\[4pt]
\end{tabular}
\caption{Contour plot of the relative error of the parameter $\gamma$ and $\beta$  measured through the probability of detection  Eq.\thinspace(\ref{SingleVacuumStateProbability}), in terms of the squeezing parameter $r$  and the uncertainty of the measurement of the coincidence operator $S$  Eq.\thinspace(\ref{CoincOperator}). We assume that in both HOM array configurations, the uncertainty in the probability of detection is the same. In the array configuration, $A_1$ is the area for the first array configuration, in which the detection of the first time delay is done. The second array has area $A_2 = \eta A_1$, with $\eta=1/4$ and $A_1 = 8 \times 10^3\,[\rm km^2]$.}
\label{CPRelErrorgammabetaSquezingPlot1rDeltaS}
\end{figure*}

In order to find other areas of the arrays, in Fig.\thinspace\ref{CPRelErrorgammabetatauPlotSqueezingr1Area} we plot a contour plot of the relative error of the parameters $\gamma$ and $\beta$ in terms of the area of the array. For simplicity, we again consider that the squeezing parameter is equal to $r$ for both configurations, and we assume the second area of the array to be a quarter of the first. As we can see, for a larger area the relative error for both parameters is constant if we vary the squeezing parameter, in this same case, $\delta \gamma/\gamma$ and $\delta \beta/\beta$ are $\sim 10^{-10}$. In this analysis we employed an uncertainty of the measurement of $S$ of $\delta S = 10^{-13}$. In order to find the uncertainty necessary to obtain the current relative errors of $\gamma$ and $\beta$, we simulate the uncertainty $\delta S$ in terms of the area and squeezing parameter. For the case in which $r$ is fixed and the area is optimized (see Fig.\thinspace \ref{CPRelErrorgammabetaSquezingPlot1AreaDeltaS}), the uncertainty $\delta S$ is $\sim 10^{-8}$ for $\gamma$ and $\beta$. On the other hand, when the area is fixed but $r$ varies (see Fig.\thinspace\ref{CPRelErrorgammabetaSquezingPlot1rDeltaS}) we need an uncertainty $\delta S \sim 10^{-7}$ with an squeezing parameter $r=1$ and for an area $A_{\rm max} = 8\times 10^{3}\;[Km^2]$.

\begin{figure*}
\centering
\begin{tabular}{c}
\includegraphics[width=90mm]{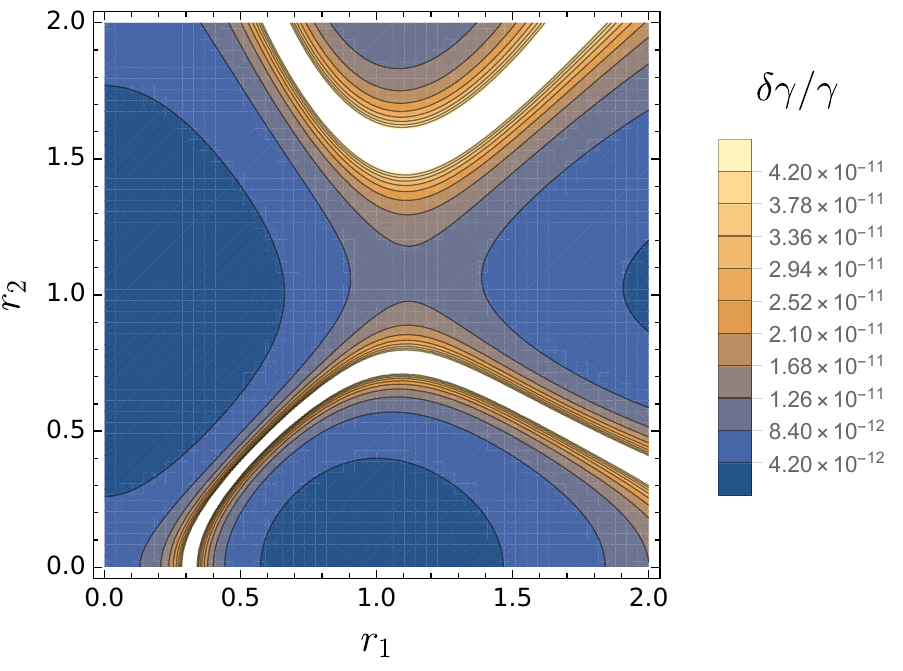} \\
(a) \\
 \includegraphics[width=90mm]{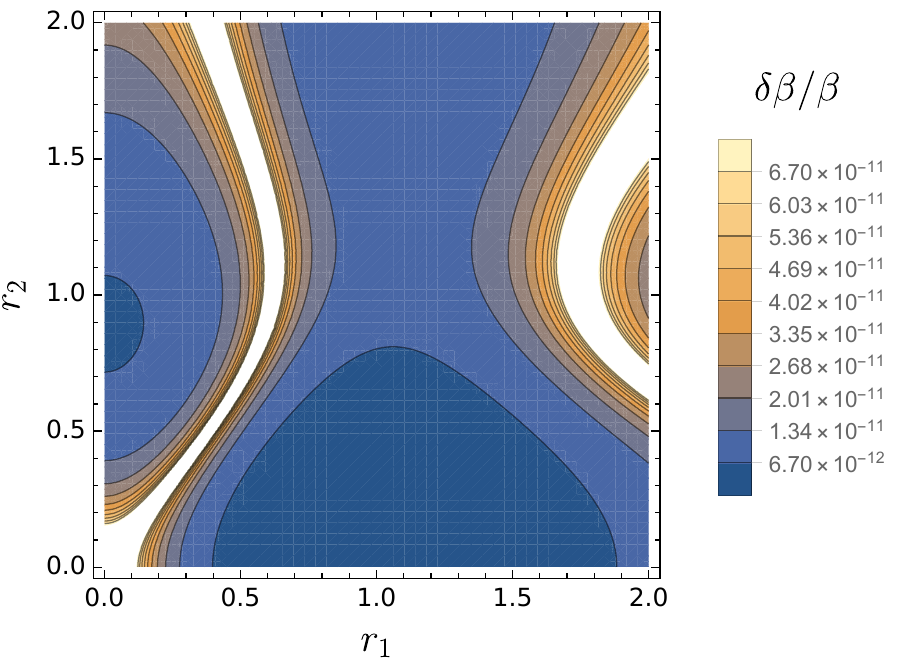} \\
(b)  \\[4pt]
\end{tabular}
\caption{Contour plot of the relative error of the parameters $\gamma$ and $\beta$ measured through the probability of detection  Eq.\thinspace(\ref{SingleVacuumStateProbability}), in terms of the squeezing parameters $r_1$ and $r_2$, employed in the first and second configuration of the HOM array, respectively. (a) Relative error parameter $\gamma$. (b) Relative error parameter $\beta$. The uncertainty in the measure of the probability of detection is $\delta S = 10^{-13}$. We assume that, in both HOM array configurations, the uncertainty in the probability of detection is the same.In the array configuration, $A_1$ is the area for the first array configuration, in which the detection of the first time delay is done. The second array has area $A_2 = \eta A_1$, with $\eta=1/4$. The mean wavelength of both photons is $\lambda = 0.995 \;[\mu m]$ and a bandwidth $\delta \lambda = 0.034\;[\mu m]$. }
\label{CPRelErrorgammabetatauPlotSqueezingr1r2}
\end{figure*}

\section{Summary and conclusions} \label{SEC6}

We have studied the problem of estimating values of the parameters $\gamma$ and $\beta$ of the post-Newtonian expansion. For this purpose, we have considered a basic setup consisting of a HOM interferometer whose arms are at different gravitational potentials. We consider the measurement of two different arrival times in order to obtain an estimation of the PN parameters $\gamma$ and $\beta$. Moreover, we study the case in which ones measures the coincident detection probability. We show that the latter method leads to an improvement in relative errors when compared with classical approaches \cite{Clifford-2014-review-on-experimental-tests-of-GR}. We consider two-photon separable states and two-mode squeezed-vacuum states to estimate the values of $\gamma$ and $\beta$.  Table \ref{Results1} shows that relative errors obtained for $\gamma$ and $\beta$ are approximately of the same order of magnitude if we consider two-mode squeezed-vacuum states. Moreover, employing this state we obtain a lower uncertainty in the estimation of the PN parameters even with a higher uncertainty in the measurement of the coincidence operator (see table \ref{Results2}). We emphasize that these schemes rely on the use of specific quantum states of light, which impose conditions on the sources employed to generate the quantum states. Fortunately, these sources are already available and are known as ultra-broadband SPDC sources \cite{Vanselow-2019-SPDC}.

One of the advantages of the  protocol discussed here   is its versatility in terms of the initial states that enter the interferometer. To improve the detection and sensitivity we can employ NOON states or cat states, among many others. An obvious modification of this scheme could be the use of a Mach-Zehnder interferometer and measuring  a different observable, instead the coincidence detection observable,  as for example the relative number of photons arriving to the output ports \cite{Pezze_2008-Mach-Zechnder-Interferometry-with-coherent-and-squeezed-vacuum-light}. A more realistic implementation should consider loss in the detection \cite{Kish-2016-Estimation-Spacetime-Parameters-in-lossy-environment} or the probability of no detection in the HOM array \cite{Chen-2019-HOM-on-a-biphoton-beat-note,Lyons-2018-attosecond-resolution-HOM-interferometry}, and include the relative velocity of the satellites \cite{Rideout-2012-Fundamental-quantum-optics-experiments-conceivable-with-satellites-reaching-relativistic-distances-and-velocities,Terno-2020-Large-scale-optical-interferometry-in-general-spacetimes}.

\begin{table}[H]
        \centering
        \begin{tabular}{||c|c|c|c||}
        \hline
           Method & $\delta \gamma/\gamma$ & $\delta \beta/\beta$ & $\delta S$  \\ \hline
          Classical tests  & $\sim 10^{-5}$ & $\sim 10^{-5}$  & experiment dependent \\ \hline
            Measurement of $\Delta \tau_i$ ($i=1,2$) &  $\sim 10^{-2}$ & $\sim 10^{-11}$ & $ \sim10^{-18}$ \\ \hline 
          Separable bipartite state &   $\sim 10^{-16}$ & $\sim 10^{-7}$ & $\sim10^{-18}$ \\ \hline 
          Two-mode squeezing (r=1) &  $\sim 10^{-10}$ & $\sim 10^{-11}$ & $\sim10^{-13}$ \\ \hline 
                  \end{tabular}
        \caption{Relative errors for $\gamma$ and $\beta$ given an area of the first array  $A_{\rm max}=8\times 10^3[km^2]$}
        \label{Results1}
    \end{table}

\begin{table}[H]
 \begin{center}
  \begin{tabular}{||c|c|c||} \hline
  	 Method & $\delta S$ for $\gamma$ & $\delta S$ for $\beta$ \\ \hline
            Measurement $\Delta \tau_i$  ($i=1,2$) & $\sim 10^{-20}$ & $ \sim 10^{-15}$ \\ \hline 
            separable biparite state & $\sim10^{-10}$ & $\sim10^{-16}$ \\ \hline
            two-mode squeezing ($r=1$) & $\sim 10^{-8}$ & $\sim 10^{-8}$ \\ \hline 
  \end{tabular}
 \end{center}
 \caption{Uncertainty necessary to achieve the current uncertainties of $\gamma$ and $\beta$}
 \label{Results2}
\end{table}

\begin{appendices}
\section{Two mode vacuum states}\label{ApSingleVacuumState}
A two mode vacuum state is given by the Eq.\thinspace(\ref{SingleVacuumState}). The output state is given by 
\begin{eqnarray}
\vert \zeta_{\rm out} \rangle &=& \frac{1}{\cosh(r)} \iint d\omega_1 \omega_2 f(\omega_1,\omega_2) \sum_{n=0}^{\infty} (-1)^n e^{in\theta}\left( \tanh(r)\right)^n \left[\frac{c^{\dagger}_{\omega_1}}{\sqrt{2}} + \frac{d^{\dagger}_{\omega_1}}{\sqrt{2}} \right]^n e^{-i\omega_1 \Delta \tau_{\gamma_1}} \nonumber \\
&& \left[\frac{c^{\dagger}_{\omega_2}}{\sqrt{2}} - \frac{d^{\dagger}_{\omega_2}}{\sqrt{2}} \right]^n e^{-i\omega_2 \Delta \tau_{\gamma_2}} \vert 0 \rangle_{12}, \nonumber \\
\end{eqnarray}
where $\Delta \tau_{\gamma_1}$ and $\Delta \tau_{\gamma_2}$, are the time delay for the paths $\gamma_1$ and $\gamma_2$, respectively.  In order to simplify the probability of detection we consider $n=1$ squeezing states, under this consideration the probability to detect one photon is given by
\begin{eqnarray}
P_{\rm cd} &=& \langle \zeta_{\rm out} \vert \hat{M}_c \otimes \hat{M}_d \vert \zeta_{\rm out} \rangle,
\end{eqnarray}
where $\hat{M}_c = \int d\omega c^{\dagger}_{\omega} \vert 0 \rangle \langle 0 \vert c_{\omega}$ and $\hat{M}_d = \int d\omega' d^{\dagger}_{\omega'} \vert 0 \rangle \langle 0 \vert d_{\omega'}$.  The amplitude of probability for $n=1$ reads
\begin{eqnarray}
\langle 1 \vert \langle 1 \vert \zeta_{\rm out} \rangle &=&
\frac{1}{\cosh(r)} \iint d\omega_1 d\omega_2 f(\omega_1,\omega_2)^{\ast} \nonumber \\
&& \langle 0 \vert c_{\omega}d_{\omega'} \left[ e^{-i\omega_1 \Delta \tau_{\gamma_1}} e^{-i\omega_1 \Delta \tau_{\gamma_2}} \vert 0 \rangle \vert 0 \rangle - e^{i\theta} \frac{\tanh(r)}{2} \left( c^{\dagger}_{\omega_1} + d^{\dagger}_{\omega_1} \right) \ \left( c^{\dagger}_{\omega_2} - d^{\dagger}_{\omega_2} \right) \vert 0 \rangle \right] \nonumber \\
& = & -\iint d\omega_1 d\omega_2 f(\omega_1,\omega_2)^{\ast} \frac{e^{i\theta}\tanh(r)}{2\cosh(r)} \langle 0 \vert c_{\omega} d_{\omega'}\left( c^{\dagger}_{\omega_1} + d^{\dagger}_{\omega_1} \right) \left( c^{\dagger}_{\omega_2} - d^{\dagger}_{\omega_2} \right)\vert 0 \rangle . \nonumber \\
\end{eqnarray}
The non-vanishing terms in the operator are the following
\begin{eqnarray}
\langle 0 \vert -c_{\omega} c^{\dagger}_{\omega_1} d_{\omega'}d^{\dagger}_{\omega'}d^{\dagger}_{\omega_2} +  c_{\omega} c^{\dagger}_{\omega_2} d_{\omega'}d^{\dagger}_{\omega'}d^{\dagger}_{\omega_1} \vert 0 \rangle &=& -\delta(\omega-\omega_1)\delta(\omega'-\omega_2) + \delta(\omega-\omega_2)\delta(\omega'-\omega_1). \nonumber \\
\end{eqnarray}
Therefore, the probability of detection, in case of $f(\omega_1,\omega_2) = f(\omega_1)f(\omega_2)$, with $f(\omega_i)$ a Gaussian distribution centred in $\omega_i$, reads
\begin{eqnarray}
P_{\rm cd} &=& \frac{1}{2}\left(\frac{\tanh(r)}{\cosh(r)}\right)^2 \left[ 1 - \exp\left(-\sigma^2 \Delta\tau^2/2 \right) \right],
\end{eqnarray}
where $\sigma$ is the spectral width of the Gaussian distribution, assumed to be equal for both photons. 

\end{appendices}
\section*{Acknowledgements}
This work was supported by the Millennium Institute for Research in Optics (MIRO) and FONDECYT Grant 1180558. M.R.-T. acknowledges support by CONICYT scholarship 21150323.
\printbibliography[heading=bibintoc]	
\end{document}